\documentclass[draftcls,peerreview]{IEEEtran}

\usepackage{cite}      

\usepackage{graphicx}  
\begin{document}
%
\title{Process Optimization and Downscaling of a Single Electron Single Dot  Memory}
%
%
\author{Christophe~Krzeminski,
        Xiaohui~Tang,
        Nicolas~Reckinger,
        Vincent~Bayot,
        Emmanuel~Dubois}

\thanks{}
\maketitle

\begin{abstract}
This paper presents  the process optimization  of a single electron  nano-flash electron memory. Self aligned single dot memory structures have been fabricated using  a wet anisotropic oxidation of a silicon nano-wires. One of the main issue  was  to  clarify  the process conditions for the dot formation. Based on the process modeling, the influence of various parameters (oxidation temperature, nano-wire shape) have been  investigated. The  necessity  of a sharp compromise between these different parameters to ensure the presence of the memory dot has been established. In order  to propose an aggressive memory cell,  the downscaling of the device has been carefully studied. Scaling rules show that the size of the original device could be reduced by a factor two. This point has been previously confirmed by the  realization of  single electron memory devices.
\end{abstract}


\section{Introduction}

\PARstart{C}{onventional} flash memory devices are believed to be  hardly scalable since the design space vanishes  below the 40 nm technology node \cite{ITRS,Kim05,Lu08}. Since a few  years, most of the area reduction has been achieved by the downscaling of isolation  and interconnection regions \cite{Deblauwe, Fazio5}. Alternative and innovative storage structures for nonvolatile memories are strongly needed as the number of stored electrons is declining to dangerous small numbers. Many emerging research devices are under investigation using  a large diversity of materials and approaches:  phase memory changes \cite{lankhorst05}, polymer memory \cite{Ouyang04},  nanocrystal memory \cite{Tiwari96}. Among all, Phase Change  nanocrystal Memory (PCM) is now acknowledged to be the leading candidate \cite{Hudgens08}.  However silicon nanocrystal remains an interesting  approach  due  to a close compatibility with some standard industrial process and  for 10 nm devices \cite{Lammers08}. In this paper, the fabrication of single electron nanoflash memories with silicon nanocrystal is under investigation. Numerous  studies have been published on  multinanocrystal memories generated  by  low-pressure chemical vapor deposition \cite{Baron03, Lombardo04} by implantation and annealing step \cite{Bonafos05} or by aerosol deposition techniques \cite{Deblauwe00}. One  strong issue faced by the multi-dots approach is the impact of the dot distribution fluctuations on the electrical properties \cite{Pernolia03}. A  relatively different approach is followed here since the idea is to  improve and investigate the limits of a single dot memory.  The fabrication of this alternative  memory is based on an original combination of classical microelectronics process steps. The main idea  is to separate a silicon dot from the channel using  anisotropic oxidation. It has already been shown that pattern dependent oxidation (PADOX) could be used to  fabricate small silicon single-electron transistors (SETs) \cite{ono}. Self-limited oxidation effects and strain have been carefully engineered to generate silicon nanostructures embedded in silicon dioxide using pattern conversion \cite{uematsu}. A different approach has been undertaken in the present work. The anisotropic character is generated using the strong dependence of the oxidation kinetics on  arsenic doping profile \cite{hoplum,hoplum2}. One issue is  to improve  the understanding of the conditions for the dot formation. Based on that background, the standard  process is first summarized in section II. Process simulation  and models limitations are discussed in section III. Section IV reports on the conditions for the dot formation   and on the influence of the oxidation temperature and of the nanowire width. Device downscaling is explored in the last section.

\pubidadjcol

\section{Device architecture and processing}

\begin{figure}[tbp]
\center
\includegraphics[width=4.3cm ,angle=-90]{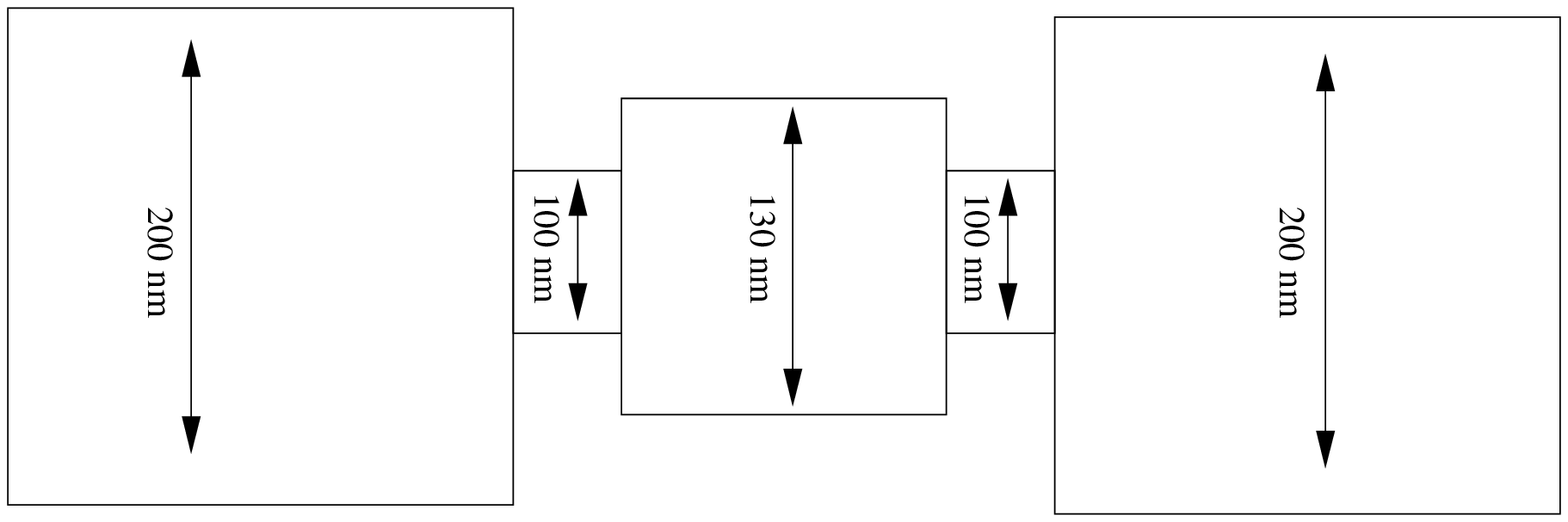}
\caption{Layout of the nanodevice. In order to generate the structure made of a single silicon dot in a channel separated  by oxidation from the source/drain contact, different regions have to defined by a lithography step. The central part of the nanodevice where a single  memory  silicon dot separated from the channel is implemented  is defined  by  a square area  of 130 nm$^{2}$. This central part is connected to thin constriction regions (with less volume to oxidize)  promoting the isolation  of the dot from the source/drain regions. Finally, larger source/drain contact regions are fabricated in order to control the transport in the channel as  a classical FET.\label{fig:caption}}
\end{figure}

\begin{figure}[tbp]
\center
\includegraphics[width=8cm,angle=0]{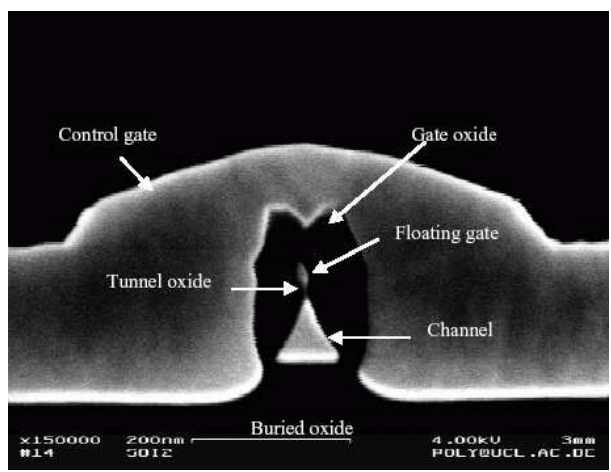}
\caption{SEM photography of the central part of the  nanoflash memory. \label{fig:mrs01}}
\end{figure}

This section briefly recalls the main process steps since  a detailed overview  can be found in  reference \cite{Xiaohui00}.  The  critical step  corresponds to the implantation through a thin SiO$_{2}$ film of a high arsenic dose (1$\times$10$^{15}$cm$^{2}$). The main objective is  to create a sharply localized and a very doped area in the silicon active layer (200 nm thick) of a SOI wafer.   A  nitride layer is deposited on the substrate.  E-beam lithography and reactive ion etching are used to pattern the nanowires.  Fig. \ref{fig:caption} presents the layout of the device. The central part of the device consists of a  130 nm by 130 nm square. Two 100 nm wide constrictions  are created in order to realize the connection between the source and drain regions and the central part. The different dimensions are defined so that a dot can be created in the central part of the device while being separated from the source/drain regions.  Finally, a wet oxidation step is performed in order to create the dot. Fig. \ref{fig:mrs01} shows a SEM photography of the central part where a single dot memory is obtained. A  nano-floating gate is clearly observed on top of the channel.

\section{Process simulations models}

Based on process simulation, the aim of this study is to clarify the conditions that lead to  dot memory existence.  Two  steps have been identified to play a key role: the wet oxidation step and the lateral overetching effect during patterning. The first one is critical for the generation and the separation of the dot/channel structure.  The second one  helps indirectly to achieve the desired pattern conversion. The main objective of this section is to discuss the limitations of process simulation with these two process steps.

\subsection{Modeling of the oxidation step}

To date, the seminal model of Deal and Grove  remains the most popular approach used in process simulators to model the oxide growth \cite{deal65}. The kinetic of oxide growth  can be simply described by the following equation :

\begin{equation}
X= \frac{A}{2} \cdot \left [ \sqrt{1+\frac{t}{A^{2}/4B}}-1 \right ].
\end{equation}

The  oxide thickness $X$  is described by a linear-parabolic relationship as a function of the oxidation duration $t$. The term $\displaystyle \Big ( \frac{B}{A}  \Big ) $ characterizes the  growth rate regime:

\begin{equation}
\begin{array}{cc}
X \sim \displaystyle \frac{B}{A} t & $for$  \quad t  \ll \frac{A^{2}}{4B}\\
\label{linear}
\end{array}
\end{equation}

\noindent whereas $B$ gives the parabolic rate constant that governs the diffusion limited regime in the limit of very long oxidation time :

\begin{equation}
\begin{array}{cc}
X \sim  \sqrt{B t} & $for$ \quad t \gg \frac{A^{2}}{4B}.\\
\label{parabolic}
\end{array}
\end{equation}
 
The  diffusivity of the oxidant species and the  reaction rate k$_{Si}$ can directly be deduced from the linear parabolic constants:

\begin{equation}
\left\{
\begin{array}{cc}
\displaystyle D= B \cdot \frac{N_{1}}{2 C^{*}}\\
\\
\displaystyle k_{Si}=\frac{B}{A} \cdot \frac{N_{1}}{C^{*}}\\
\end{array}
\right.
\label{diff_stress2}
\end{equation}

\noindent where C$^{*}$ is the oxidant solubility in the oxide and N$_{1}$ is the number of oxidant molecules incorporated into a unit volume of the oxide.

The well-known breakdown of  the Deal and Grove model \cite{Massoud85a, Massoud85b, Krzeminski07}  in the ultra-thin regime is not a significant issue here since  only very long wet oxidation are considered. However, three points have to be carefully addressed : the mesh refinement, the influence and the diffusion of arsenic  on the oxidation reaction rate and the global influence of  strain  on oxidation.

\subsubsection{Grid refinement}

One critical issue is the mesh refinement since the  dot radius is less than 20 nm. The refinement must be  as dense as possible in the region of high doping and more relaxed in other parts of the system (nitride mask, buried oxide). The second aspect is that the grid refinement must be preserved during the simulation of the oxidation process. One solution  to preserve the quality of the initial grid during oxidation is to  perform simulation using very small time-steps (less than 1 ms) and regenerate and adapt the grid frequently.

\subsubsection{Influence of arsenic}

\begin{figure}[tbp]
\center
\includegraphics[width=8cm,angle=0]{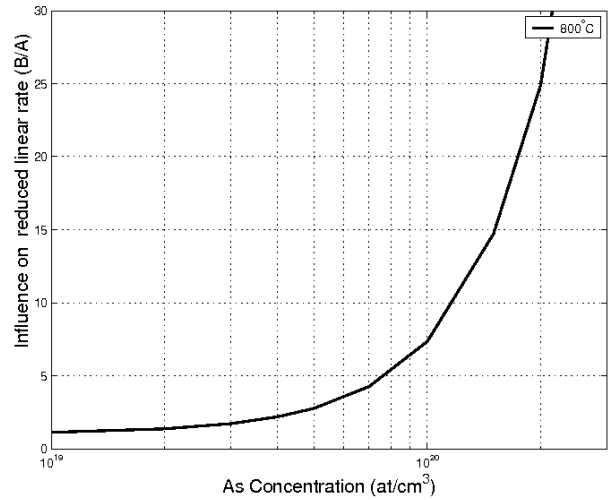}
\caption{Direct influence of the Arsenic concentration on the reduced oxidation linear rate predicted by the Ho and Plummer theory \cite{hoplum, hoplum2}. Above an arsenic concentration of  10$^{19}$ at/cm$^{3}$, $\displaystyle (B/A)^{doped} = (B/A)^{0}$ whereas an increase by a factor of 7 is observed at   10$^{20}$ at/cm$^{3}$.\label{fig:facdop}}
\end{figure}

The influence of arsenic on the reaction rate is described by the theory of  Ho and  Plummer \cite{hoplum, hoplum2}. The physical  effect behind the oxidation rate enhancement is that the  silicon Fermi level is shifted by doping. The change of the  Fermi level enhances the  concentration of vacancy that naturally provides more reaction sites for the oxidation mechanism. The linear oxidation rate in presence of dopants can be  expressed by:

\begin{equation}
\Big (\frac{B}{A} \Big )^{doped}= \Big (\frac{B}{A} \Big)^{0}\cdot  \Big [1+\gamma \cdot (C_{v}-1) \Big ] 
\label{one}
\end{equation}

\noindent where $\Big (\frac{B}{A} \Big )^{0} $ is the intrinsic linear reaction rate for wet oxidation which depends on many other parameters (temperature, pressure), $\gamma$ is an empirical temperature dependent parameter and $C_{v}$ is the normalized total vacancy concentration. Since dopants are assumed to influence only the oxidation rate, it should be noticed that the anisotropy will be maximum in the reaction limited regime (Eq. \ref{linear}) and negligible in the diffusion limited regime (Eq. \ref{parabolic}). Fig. \ref{fig:facdop} shows the influence of the linear oxidation rate as a function of the arsenic concentration predicted by the Ho and Plummer model \cite{hoplum, hoplum2}. The concentration should exceed 3$\times$10$^{19}$ at/cm$^{3}$ in order to observe an increase.

Another effect related to arsenic is the thermal diffusion during the oxidation process. A reduction of the initial arsenic gradient probably takes place during the oxidation process.   In order to describe the  diffusion  of arsenic during this step, a standard  five stream  model has been used \cite{Dunham92}. The main issue is to have a correct description of the amplitude and the shape of the arsenic peak. A correct description of the arsenic tail and TED (Transient Enhanced Diffusion) effects is not crucial here.

\subsubsection{Influence of strain}
                                                                   
In our specific case, the presence of arsenic has a major impact on the reaction rate. On the other hand,  strain  are expected to  reduce the oxidation rate. The presence of a nitride mask on top of the nanowire and the concave shape  of the nanowire generates strain during the oxidation process. In most of process simulators,  stress   can be included in the basic  Deal and Grove model by considering stress dependent parameters. As proposed by Kao et al. \cite{Kao87a, Kao87b} and Sutardja and Oldham \cite{Sutardja89}, the hydrostatic pressure $P$ is often assumed to influence the oxidation diffusivity by:

\begin{equation}
\left\{
\begin{array}{cc}
\displaystyle D^{\sigma}_{0}=D_{0}\cdot\exp \big [-\frac{P V_{d}}{k_{B}T}\big]& $for$ \quad P > 0\\
\displaystyle D^{\sigma}_{0}=D_{0}& $for$ \quad P \leq 0 \\
&\\
\end{array}
\right.
\label{diff_stress3}
\end{equation}

\noindent On the other hand, a compressive  normal stress  component at the Si/SiO$_{2}$ interface  reduces the oxidation rate:

\begin{equation}
\left\{
\begin{array}{cc}
\displaystyle \Big ( \frac{B}{A} \Big )^{\sigma}=  \Big ( \frac{B}{A} \Big )\cdot\exp \big [-\frac{\sigma_{nn}V_{k}}{k_{B}T}\big]& \quad for \quad \sigma_{nn} < 0\\
\displaystyle \Big (\frac{B}{A} \Big )^{\sigma}=  \Big ( \frac{B}{A} \Big ) \quad  $for$ \quad \sigma_{nn} \geq 0\\
\end{array}
\right.
\label{diff_stress4}
\end{equation}

\noindent where $k_{B}$ is the Boltzmann constant and $T$ is the oxidation temperature in Kelvin, $P$ is the hydrostatic pressure:

\begin{equation}
P=-\frac{1}{2}(\sigma_{xx}+\sigma_{yy})
\end{equation}

\noindent and $\sigma_{nn}$ is the stress normal to the Si/SiO$_{2}$ interface, V$_{d}$  and V$_{k}$  are the  corresponding activation volumes. A viscoelastic approach has been used in order to take into account  strain relaxation in the oxide \cite{Senez94}

\subsubsection{Model limitations}

\begin{figure}[tbp]
\center
\includegraphics[width=8cm]{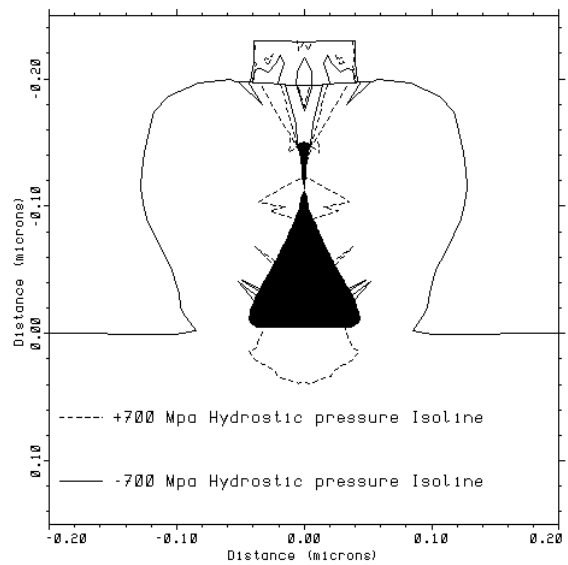}
\caption{The hydrostatic pressure  characteristic of the  strain field in the oxide  generated by the dot formation after the wet oxidation step is reported. Black regions correspond to the two silicon parts (channel and dot). Both the compressive and the tensile isoline for an hydrostatic pressure of 700 Mpa have been calculated.
\label{fig:stress_field}}
\end{figure}

As previously described, dopant diffusion and oxidation are strongly related in our specific case. The use of a simulator where all these different models are implemented is required.  Moreover, severe constraints are imposed on the grid refinement.  Both DIOS \cite{dios} and TSUPREM4 \cite{Tsuprem4} process simulators have been used. Relatively close results have been obtained as far as the kinetics of  the  dot formation is investigated. Thank to the DIOS features advanced mesh refinement  it has been possible also to investigate the oxidation of the dot.  In order to access  the evolution of the various parameters of the memory cell, a monitoring tool has been developed \cite{Krzeminski05}.  The main limitation in our study is the description of the mechanical strain field in process simulators.  Fig.  \ref{fig:stress_field} illustrates this point where the strain field in the oxide layer is reported for the simulation of the standard process described in the previous section. Two contours ($\pm$ 700 Mpa) of the hydrostatic pressure are reported based on  TSUPREM4 simulation. Very large  compressive stress levels are observed in the vicinity of the dot whereas the nitride mask is subject to a high tensile stress.  As far as we approach the outside silicon dioxide interface, stress decreases as the  relaxation is much more important. Such a large level of stress at the nanometer scale indicates  that we are too close to the limits for a  finite elements simulations to expect  a correct description of the mechanical stress around the dot.  The use of coupled atomistic/continuous models could be a solution \cite{Curtin03}. The second limitation is the use of a viscoelastic approach  to describe the  oxide deformation \cite{Senez94}. It has already been shown by Rafferty et al. \cite{Rafferty89b} that linear viscous approach strongly overestimates the stress generated during cylinder oxidation at low oxidation temperature and large deformation \footnote{A two dimensional implementation of an elastoplatic approach is very complex and subject to strong numerical issues. A rough estimation can be obtained by  the application of the Rafferty model \cite{Rafferty89b}  to our system. Considering  a silicon cylinder of 20 nm and a oxide cylinder of 50 nm gives a radial compressive strain component of 1.88 Gpa and an hydrostatic pressure of 454 Mpa at the Si/SiO$_{2}$ interface between the two cylinders}.  In principle, an  elastoplastic model would be necessary to properly estimate  the stress field surrounding the silicon dot for low oxidation temperatures \cite{Rafferty89a}. The third limitation is the modeling of   the pile-up of arsenic at the Si/SiO$_{2}$ interface \cite{Steen07}. {\it Despite these fundamental modeling restrictions,  it will be shown that interesting trends  can be extracted based on the relatively good description of experimental oxidation kinetics in the presence of arsenic by the coupling of the Ho and Plummer model \cite{hoplum,hoplum2} and the Deal and Grove models \cite{deal65}}.

\subsection{Modeling of the overetching step}

\begin{figure}[tbp]
\begin{center}
\includegraphics[width=8cm]{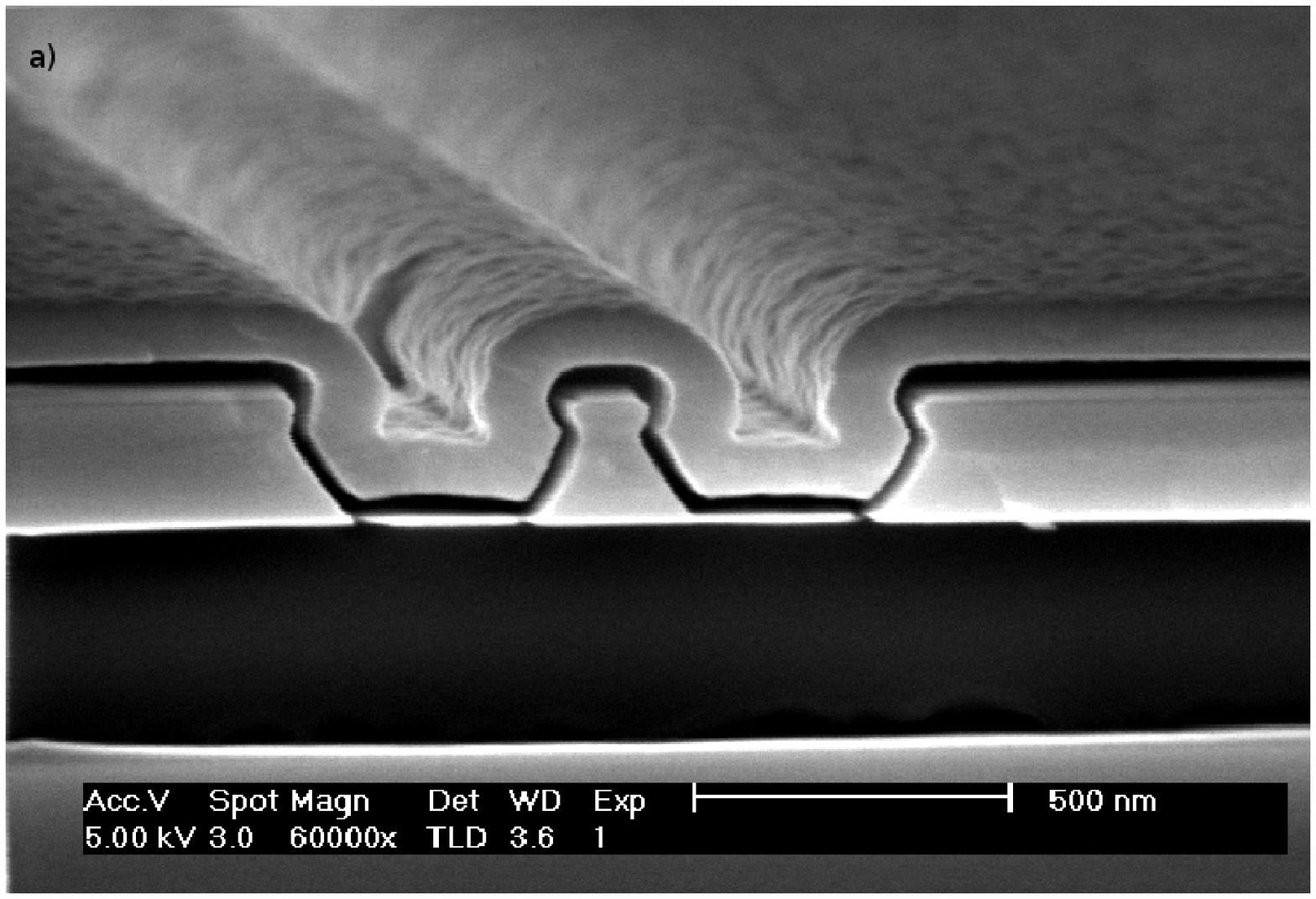}
\includegraphics[width=8cm]{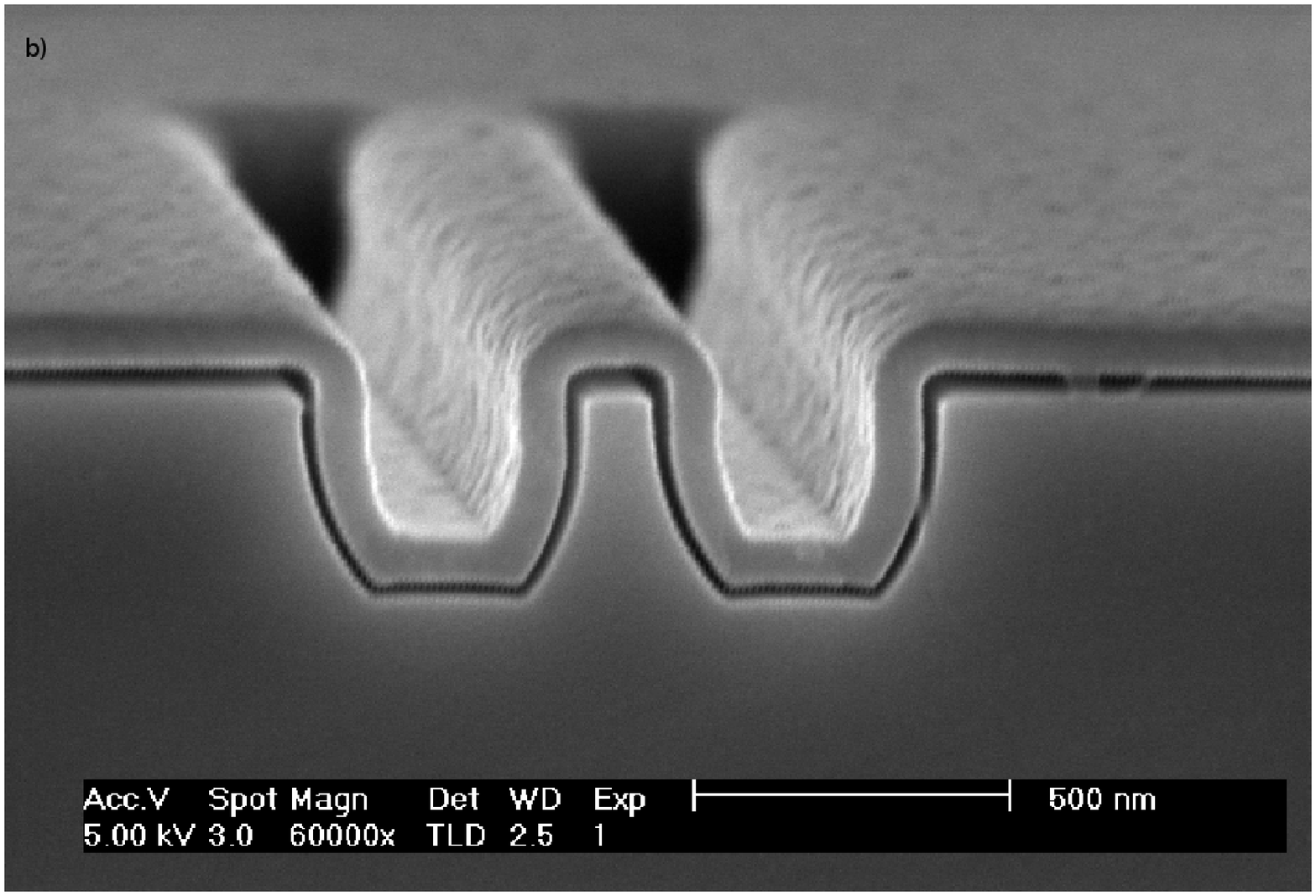}
\end{center}
\caption{Influence of the doping concentration on the shape of the mesa structure. a): The arsenic (1$\times$10$^{15}$cm$^{2}$, 70 keV) profile gives  overetching at the top of the structure. This leads to a  trapezoidal shape with a neck. b): The classical trapezoidal shape for undoped silicon.\label{fig:etch1}}
\end{figure}

\begin{figure}[tbp]
\center
\includegraphics[width=8cm]{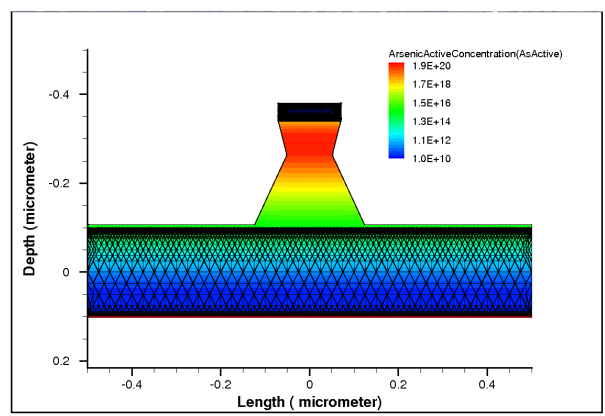}
\caption{The simulated counterpart for a 130 nm configuration. The arsenic gradient in the silicon layer is shown.\label{fig:etch2}}
\end{figure}

Due to the strong concentration of dopants, it has been observed experimentally that the top of the mesa structure is overetched during its creation. This overetching effect has been observed and explained by Winters et al. \cite{Winters}. The concentration of negative etching ions  is increased due to the electrostatic  interaction between  fluorine ions from the reactive plasma at the surface and the dopant charges in the silicon. The increase of the etching rate in the high doping region leads to a  trapezoidal shape for the  mesa-structure with a neck at the top concentration.  This effect has a major impact experimentally on the final device structure and facilitates the creation of the dot. The SEM image (Fig. \ref{fig:etch1}) clearly shows the presence of overetching in the presence of  strong arsenic doping. This effect must be discussed and incorporated in our two dimensional  simulations. An empirical approach  has been adopted in order to match as finely as possible the experimental shape.  Fig. \ref{fig:etch2} shows the shape of the mesa-structure matching the experimental configuration of Fig 5.b). The neck at the top is approximated by a simple trapezoidal shape. Three etching steps are used in the process simulator to generate the full structure. The nitride mask etching is simulated by a perfect anisotropic etching. Next 90 nm of the silicon layer are etched with an angle of 104$^{\circ}$ with respect to the horizontal direction to simulate the overetching and the presence of the neck. Finally, a classical anisotropic etch with an angle of 60$^{\circ}$ (from the horizontal direction) is performed to match the shape of the mesa-structure at the bottom. All of this process leads to an accurate shape of the mesa-structure matching the experimental one.

\section{Conditions of the dot formation}

Experimentally, the creation of a  silicon dot  has always been observed for  80 minutes of wet oxidation at 800$^{\circ}$C.  Process variations  show that the oxidation  duration should be  defined very finely in order to completely liberate the dot to avoid a complete consumption by the oxidation mechanism.  Based on process simulation, the objective is therefore  to investigate the best process conditions for dot formation. Two parameters have been identified for their influence on the dot formation: the oxidation temperature and the nanowire geometry.

\subsection{Influence of the oxidation temperature}

\subsubsection{Lifetime of the silicon dot}

\begin{figure}[tbp]
\center
\includegraphics[width=7cm]{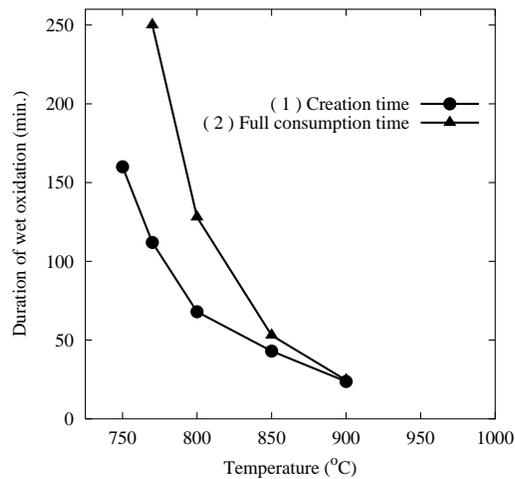}
\caption{Evolution of the times of oxidation to (1) create and totally  (2) consume  the silicon dot, for the two configurations,  as a function of the oxidation temperature.
\label{fig:time}}
\end{figure}

The most important factor for the reliability of our device is the time window during  which the dot is formed but not yet consumed. This point is governed by the difference between the oxidation time to create the dot and that to consume it. Fig. \ref{fig:time} reports the evolution of these parameters as a function of the oxidation temperature  for various thermal budgets. First of all,  an hyperbolic decrease ($\propto$ t$^{-1}$) is observed for both the creation and consumption time as a function of oxidation temperature. For example, the creation time is 160 min. at 750$^{\circ}$C and  reduces to 24 min. at 900$^{\circ}$C. This result tells us that the oxidation temperature must be higher than 750$^{\circ}$C to keep a reasonable processing time.  On the other hand,  it has been observed that the difference between the  creation time  and the consumption time vanishes rapidly with increasing temperature. At 900$^{\circ}$C, the dot is  created and consumed  at the same time, therefore the recommended wet oxidation temperature range is [775$^{\circ}$C-850$^{\circ}$C]  in order  to ensure the dot existence. The best compromise is around  800$^{\circ}$C which  ensures the presence of the dot in the central region and allows the full consumption of silicon in the constriction regions. 

\begin{figure}[tbp]
\center
\includegraphics[width=2.5in]{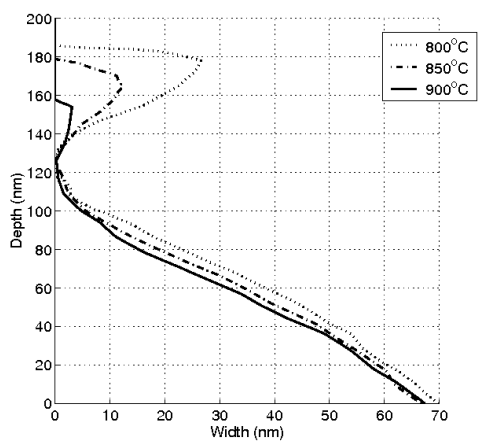}
\caption{Evolution of the mesa structure for various oxidation temperatures (800$^{\circ}$,  850$^{\circ}$,  900$^{\circ}$). For each temperature, the structure has been  reported  just at the  oxidation time  previously determinated  in Fig. \ref{fig:time}\label{fig:shape2}}
\end{figure}

The oxidation temperature has also clear impact  on the size of the dot. The evolution of the dot shape as a function of the oxidation temperature has been simulated. Fig. \ref{fig:shape2} reports both the channel and dot cross sections. It can be observed that the size of the dot is strongly reduced by  increasing the temperature.  Fig. \ref{fig:shape2} shows that the dot maximal half  width is about 25 nm at 800$^{\circ}$C and reduces to less than 5 nm for 850$^{\circ}$C. For a higher temperature than  900$^{\circ}$C, no dot is created.

\begin{figure}[tbp]
\center
\includegraphics[width=3in]{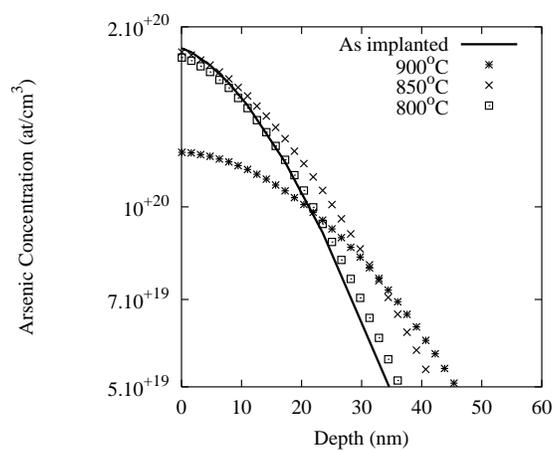}
\caption{Evolution of the arsenic profile with different thermal budgets. For each  temperature, the oxidation duration is set to correspond exactly to dot creation  (as reported in figure \ref{fig:time}).
\label{fig:prof}}
\end{figure}

Obviously, the oxidation temperature has a clear impact on the process reliability. It directly influences  the oxidation parameters such as the oxidation rate or  the strain  relaxation mechanism. However, the main striking effect is  the reduction of the arsenic gradient responsible for the dot creation, which in turn affects the dot formation. The evolution of the arsenic profile upon annealing in an inert ambience is reported. in Fig. \ref{fig:prof} for fixed temperatures and durations corresponding to the onset of the dot creation, as obtained from Fig. \ref{fig:time}. For temperature below 850$^{\circ}$C, the arsenic profile remains essentially unaffected  by thermal diffusion. Dopant diffusion  becomes significant at 900$^{\circ}$C with a strong broadening of the arsenic profile\footnote{A relevant comparison can be performed between our simulation and the experiments on arsenic TED by Solmi et al. \cite{Solmi}. It has been observed that effectively, the maximum does not diffuse at  750$^{\circ}$C and 800$^{\circ}$C. When the temperature reaches 900$^{\circ}$C, a clear reduction of the maximum level of doping is observed for a 30 min. annealing.}\footnote{ It can also be noticed that the Arsenic maximum (2 $\times$ 10$^{20}$at/cm$^{3}$) is just below the concentration where the diffusivity dramatically increases with doping concentration \cite{Larsen93}.}. It is clear that the broadening and the flattening of the arsenic profile  reduce the anisotropic character of the oxidation step and explains the difficulty to observe the dot creation at  temperatures above 850$^{\circ}$C.

\subsection{Influence of the nanowire width}

\subsubsection{Critical linewidth (experiment) }

\begin{figure}[tbp]
\begin{center}
\includegraphics[width=6cm]{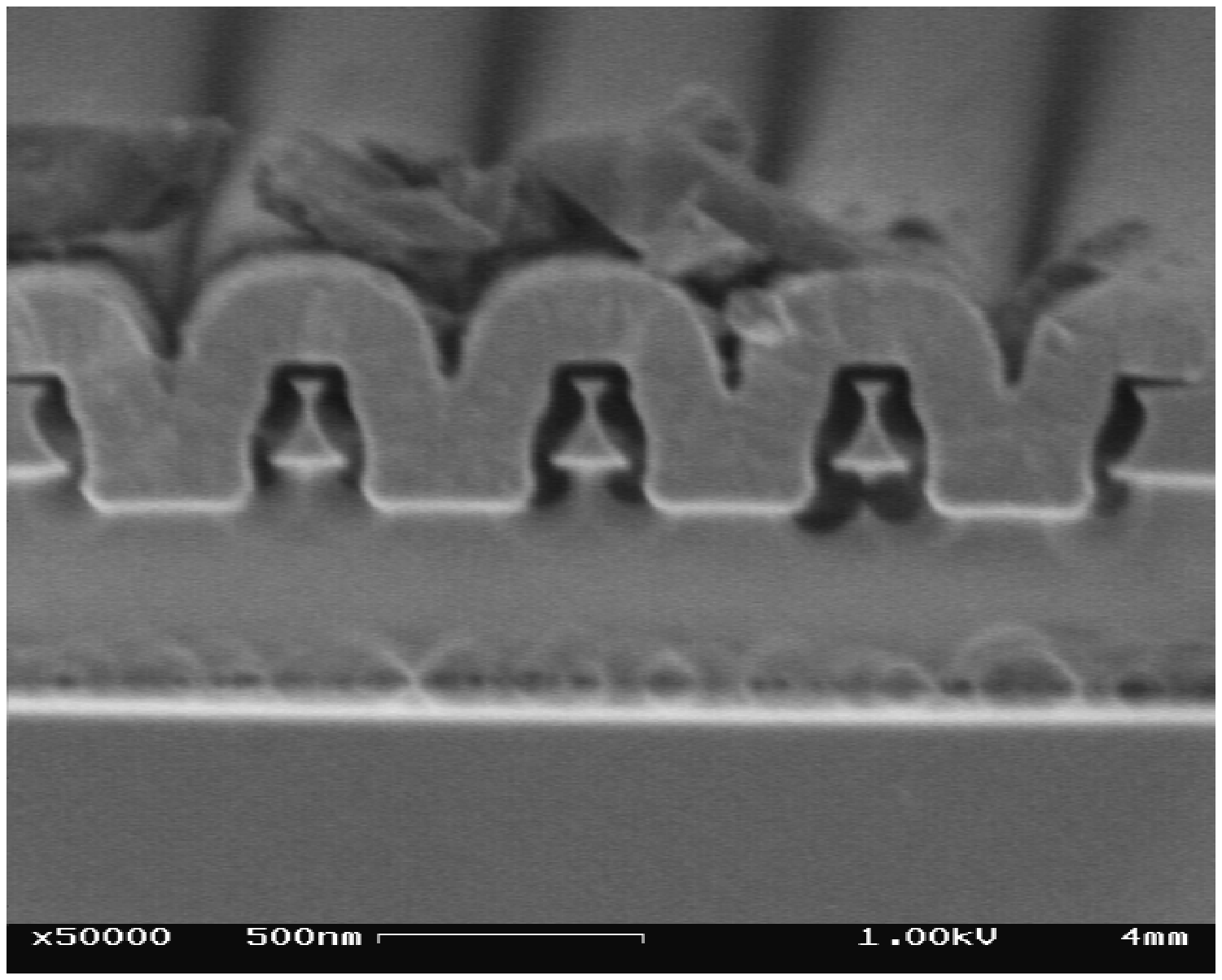}
\includegraphics[width=6cm]{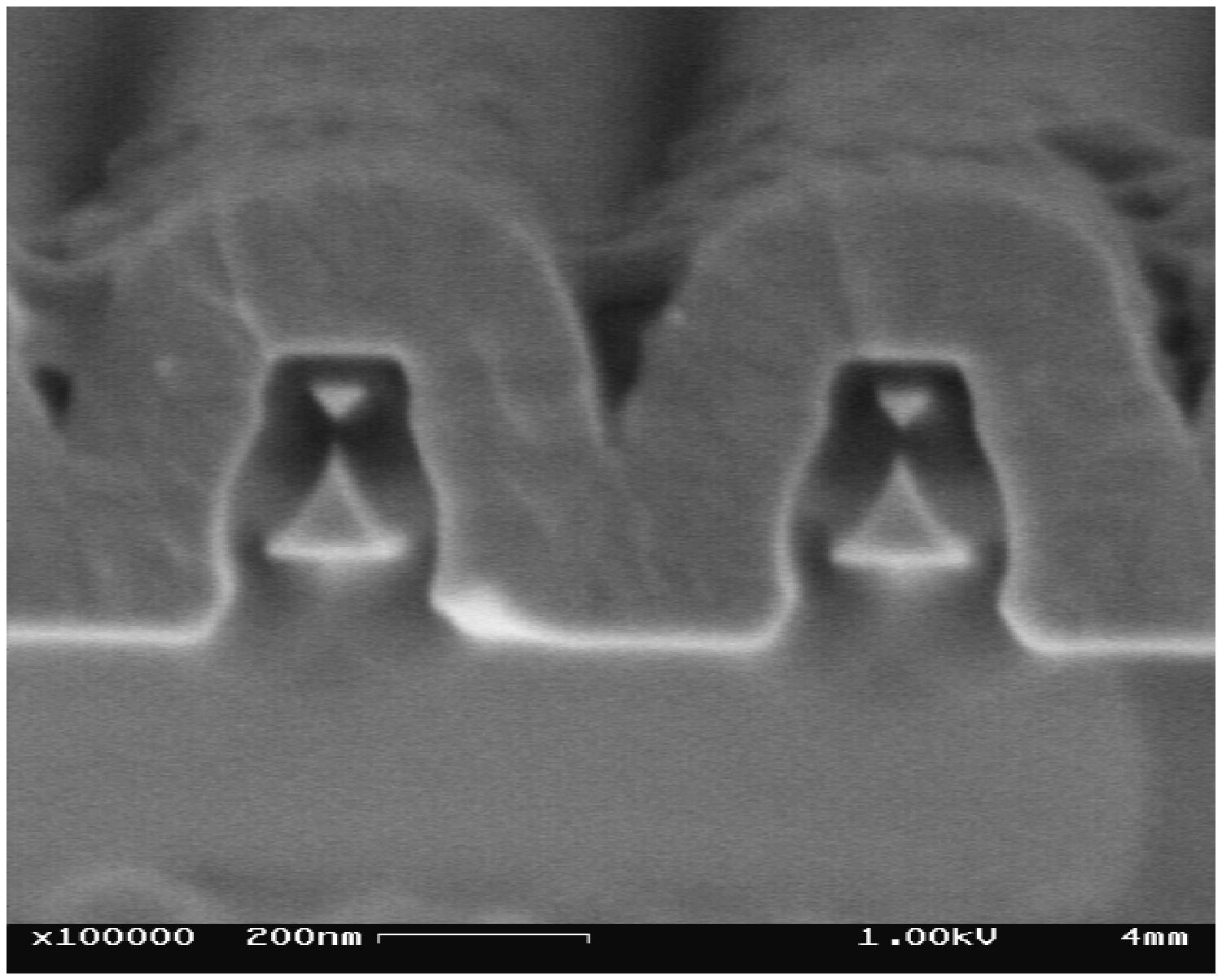}
\includegraphics[width=6cm]{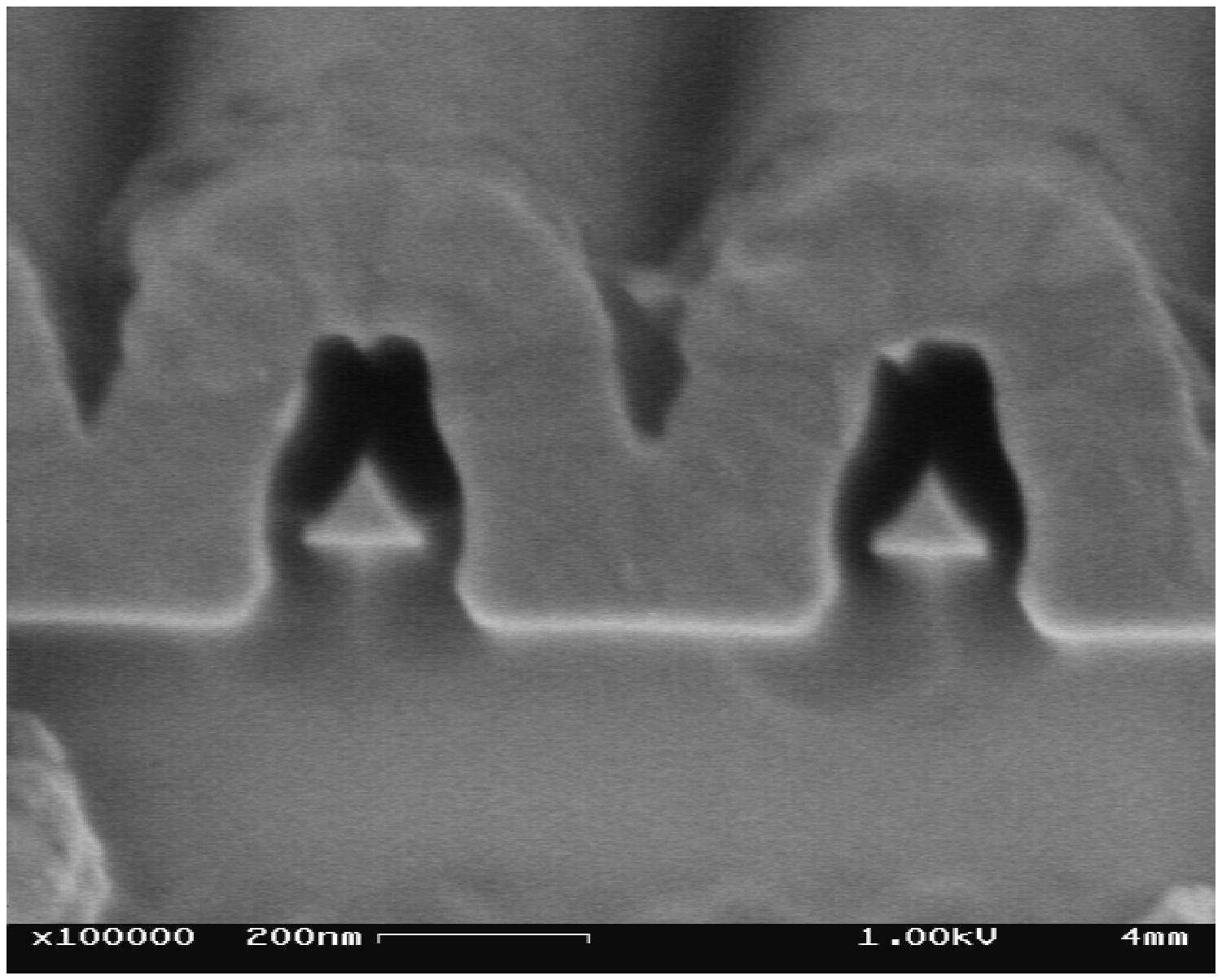}
\end{center}
\caption{Cross sections after the oxidation step  for (200 nm, 120 nm and 80 nm) line widths corresponding to the different part of the device.
\label{fig:experiment}}
\end{figure}

The main objective  is to estimate the influence of  the nanowire initial  width (at the top of the mesa-structure) on the dot existence. All the other process parameters related to the implantation or the oxidation step are those reported in section II.  In the present experimental study, the top linewidth of the mesa-structure defined by the lithography step varies from 200 nm down to 50 nm. Fig. \ref{fig:experiment} presents  some cross sections  after  the oxidation step for various linewidths (80, 120 and 200 nm). Precise data about the dot size could not be estimated from SEM analysis but the nanofloating gate diameter is about 20 nm with a significant error bar  \cite{Xiaohui06}.  The critical experimental linewidth that ensures  the dot existence for this set of experiments is observed for 120 nm $\pm$ 10 nm. The dot is not separated from the channel for a linewidth larger than 130 nm which   is therefore favorable to generate the source/drain regions. On the other hand, for a narrower linewidth than 100 nm, no dot is  observed from SEM cross section analysis  and the configuration  seems  suitable for the constriction  regions. However, it must be kept in mind that the experimental technique used to observe the dot could weaken the structure since some of the  SiO$_{2}$ cover is removed and that a very small dot could be present but not observed.

\subsubsection{Critical linewidth (modeling aspects)}

A theoretical study has been undertaken in order to quantify the influence of this parameter. The top linewidth of the mesa-structure used in simulation  varies from 200 nm down to 50 nm. The overetching effect is taken into account as described in section III.B.

\begin{figure}[tbp]
\center
\includegraphics[width=7cm]{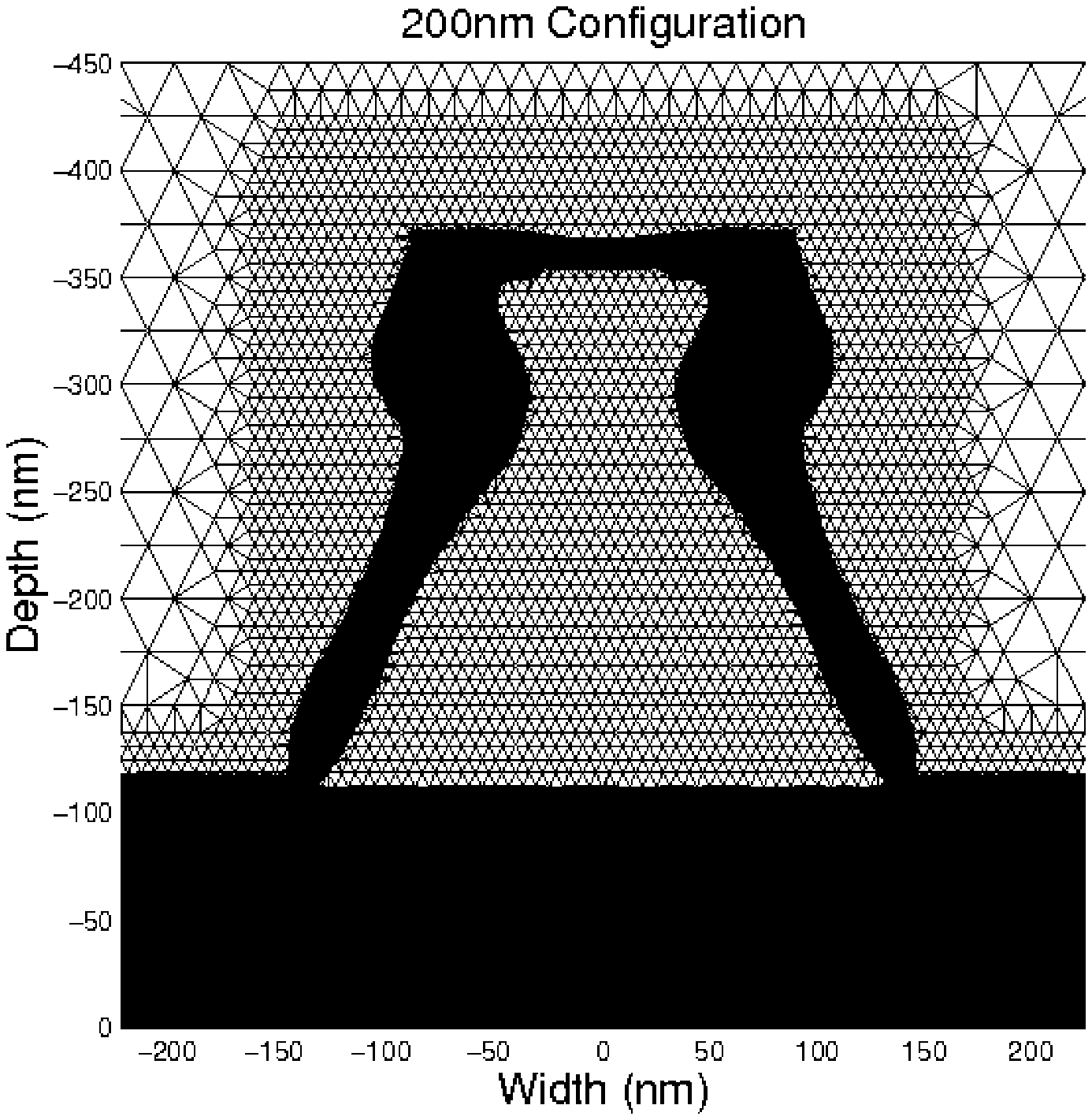}
\includegraphics[width=7cm]{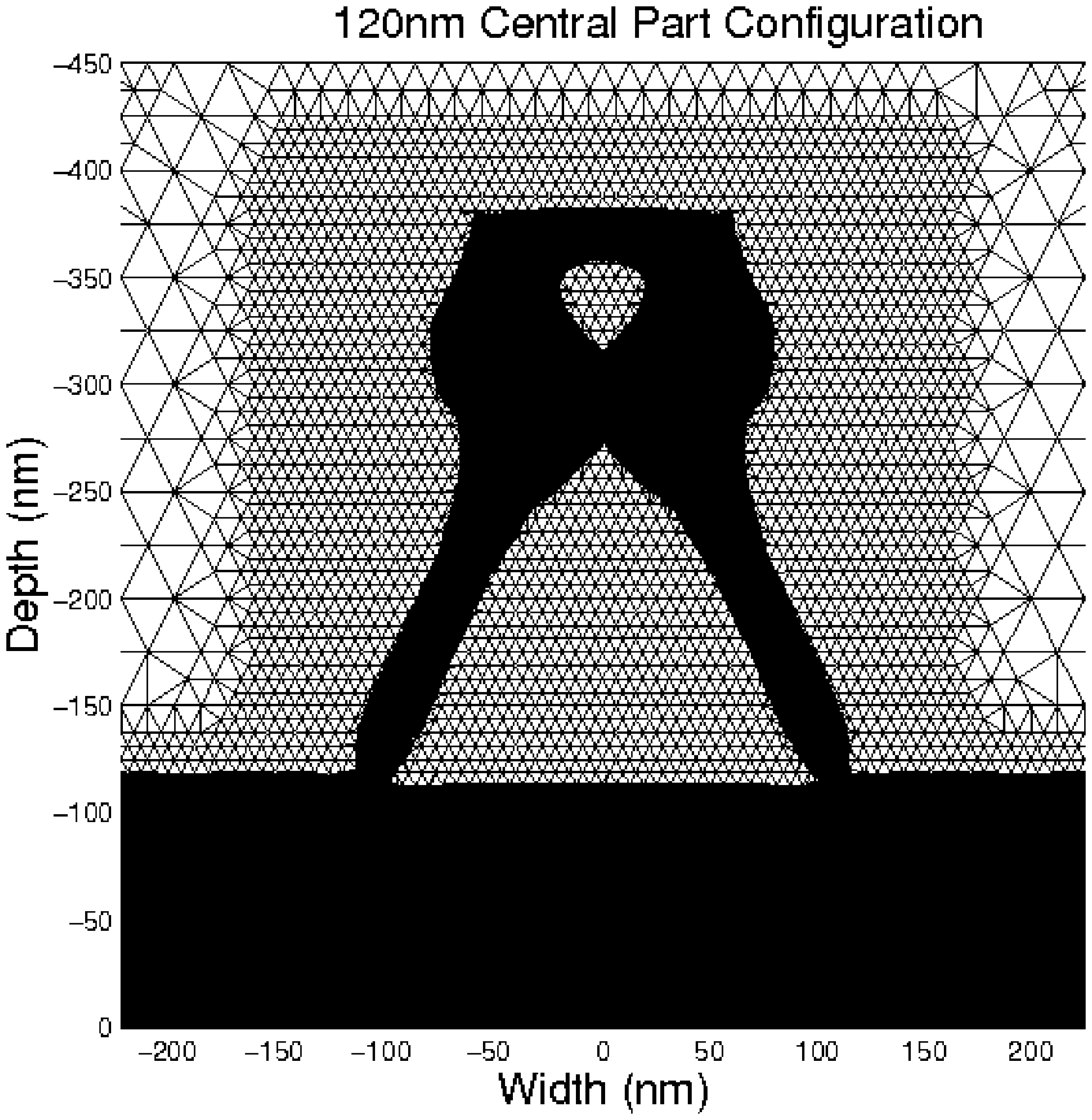}
\includegraphics[width=7cm]{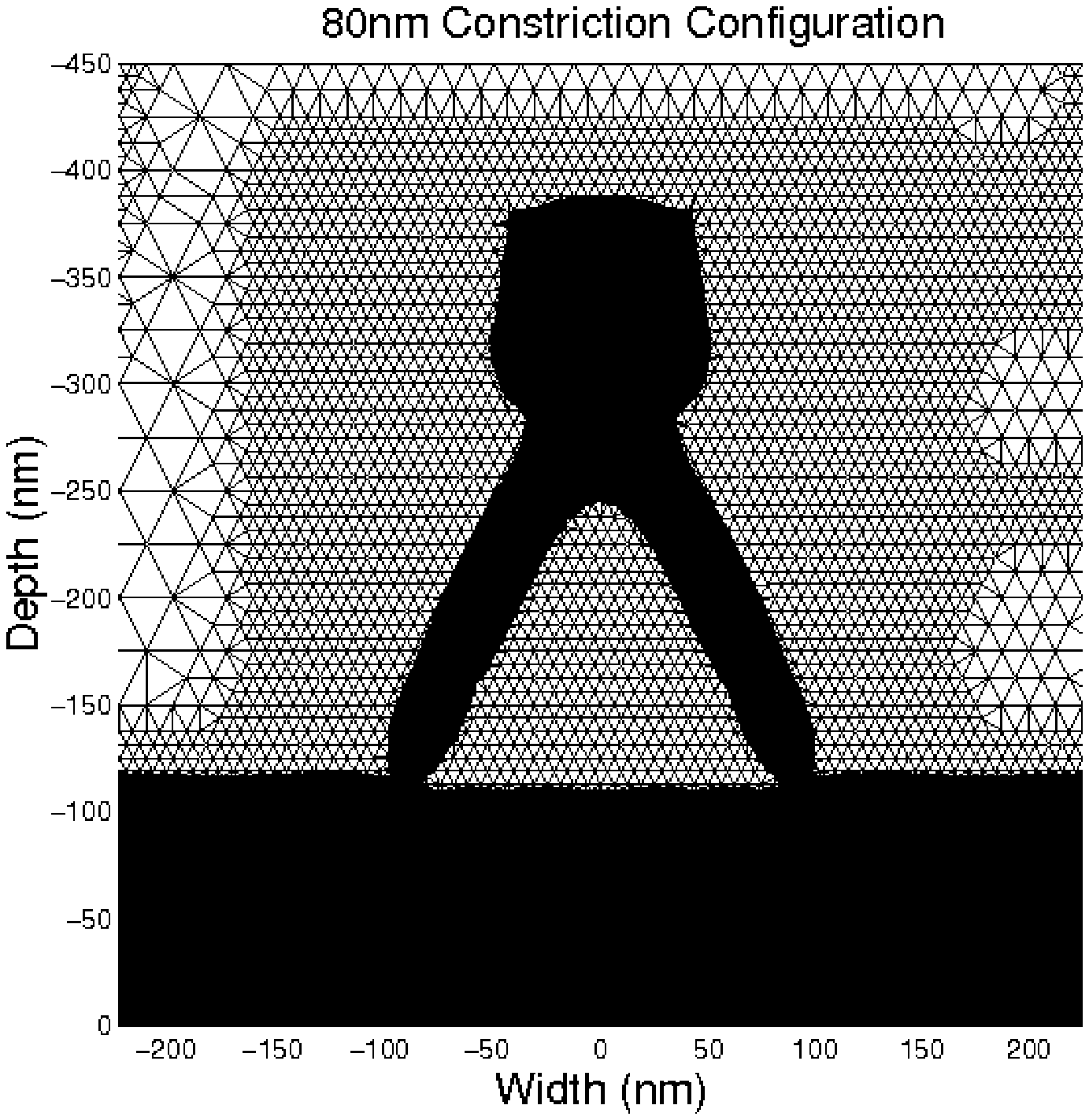}
\caption{Some configurations illustrating the critical linewidth experiments (respectively 180 nm, 120 nm, and 100 nm). All the process parameters have been kept constant (oxidation temperature 800$^{\circ}$C, duration 88 min.) and the linewidth varies from 200 nm down to 50 nm. For a linewidth larger than 130 nm, the dot is not liberated from the channel. A dot is observed  for 120 nm $\pm$ 10 nm.\label{fig:central_soi}}
\end{figure}

Fig. \ref{fig:central_soi} presents the simulation of the oxidation step for various linewidths (80, 120 and 200 nm) corresponding to the constriction, central part and source/drain regions respectively.  The direct comparison between the simulated configuration (Fig. \ref{fig:central_soi}) and their experimental counterpart (Fig. \ref{fig:experiment}) is of particular interest. It can be noticed that there is a global  agreement. Moreover,  the triangular shape of the channel as well as the dot geometry (shape, position) is well described. The dot-channel distance is estimated above 45 nm and can be compare to the experimental value of $\sim$  40nm  in Fig. \ref{fig:central_soi}.  The main  discrepancy is observed for the bottom channel since clearly  less oxidation is predicted \footnote{Since the evolution of all the other geometrical parameters are reasonably described, such a large difference could be reasonably explained reasonably  by the presence  experimentally of  a  thinner silicon after etching compare to the configuration of Fig. \ref{fig:etch1}. This layer plays a major role as it tends to delay the oxidation of the bottom part of the  channel}.  Fig. \ref{fig:param_soi} presents also a nice results  on the dot creation line as simulation predicts almost exactly the experimental linewidth where a dot is first  observed. From the simulation point of view, the dot is  just separated from the channel for a 125 nm nanowire linewidth. On the other hand, the theoretical critical  range for the dot existence is clearly much larger 40 nm compared to the experimental one. Two possible explanations can be proposed about this difference. The first one is related to the limits of the experimental technique used to observe the single dot as discussed just before-mentioned. The other one is related to the models  limitations discussed in section III.A. Stress levels with decreasing  silicon dot size  are probably overestimated which could lead to the underestimation of  wet oxidation kinetics  and a larger critical range for the dot existence.  On the other hand with respect to the experimental limits, simulation confirms that the silicon  dot is completely consumed  for a 80 nm  wide nanowire  configuration and that self-limited oxidation effects \cite{Bonafos06}  are not sufficient to stop the reaction in our present case.

\subsubsection{Simulation of  the dot characteristics}

\begin{figure}[tbp]
\center
\includegraphics[width=7cm]{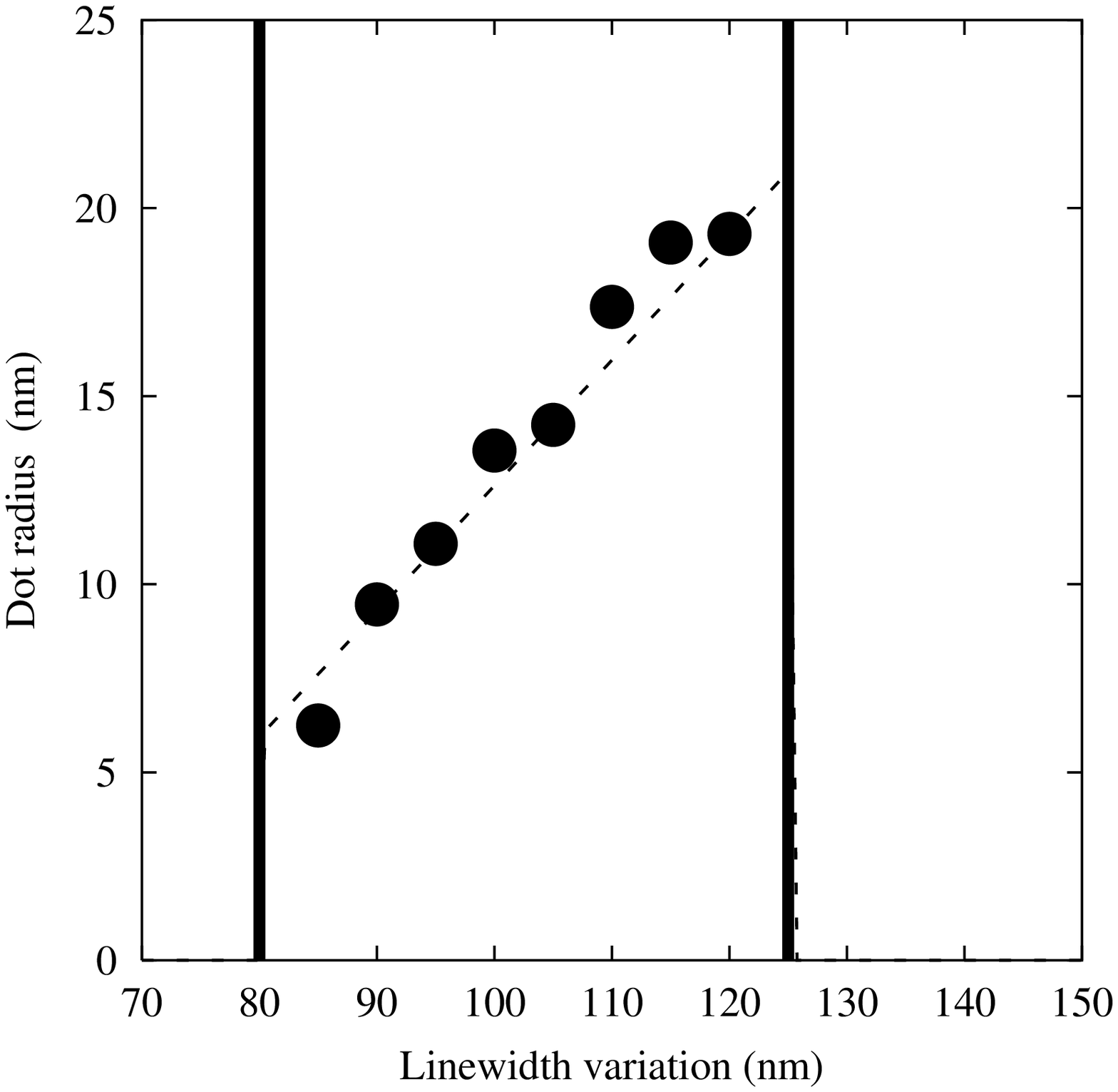}
\includegraphics[width=7cm]{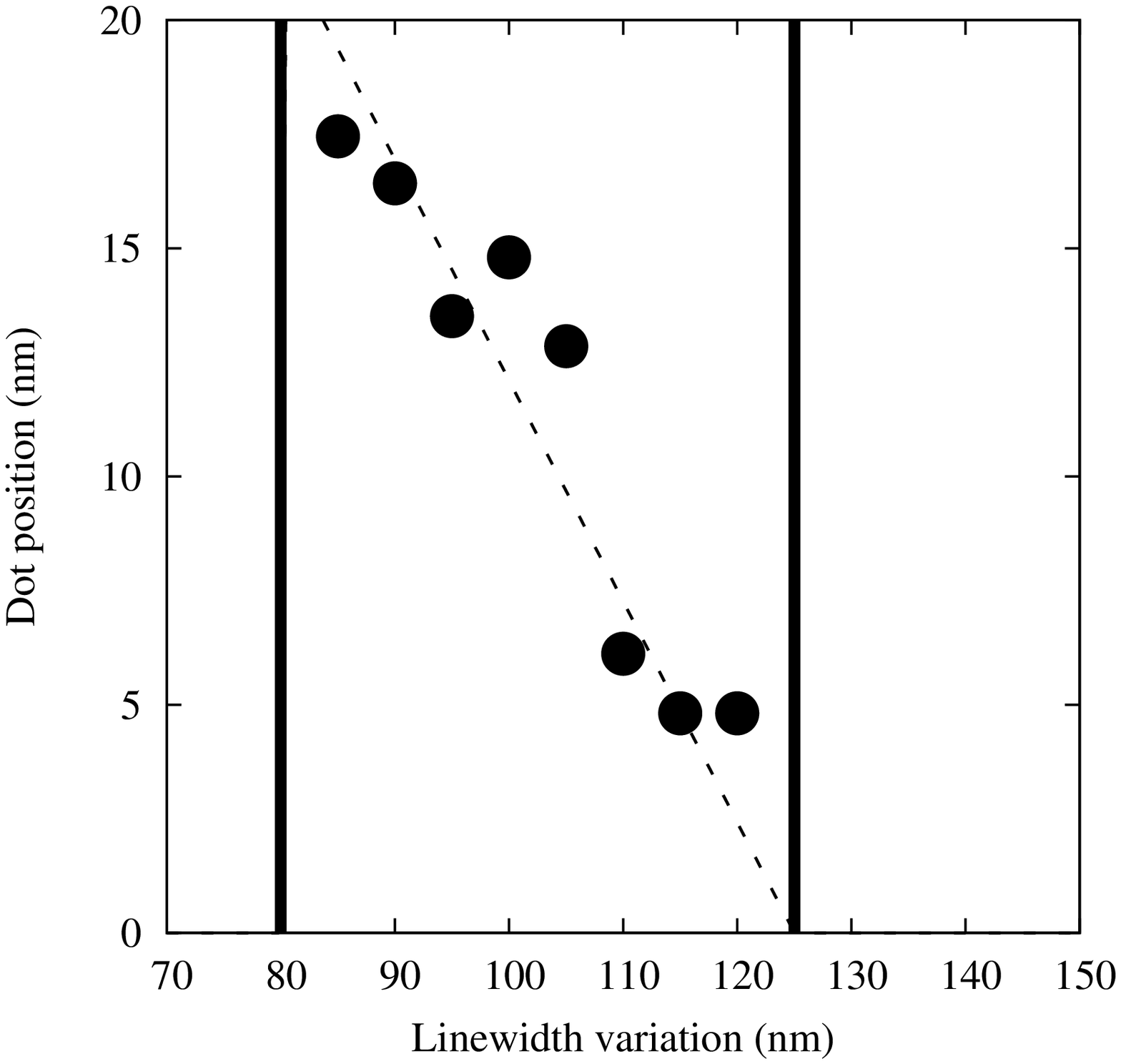}
\includegraphics[width=7cm]{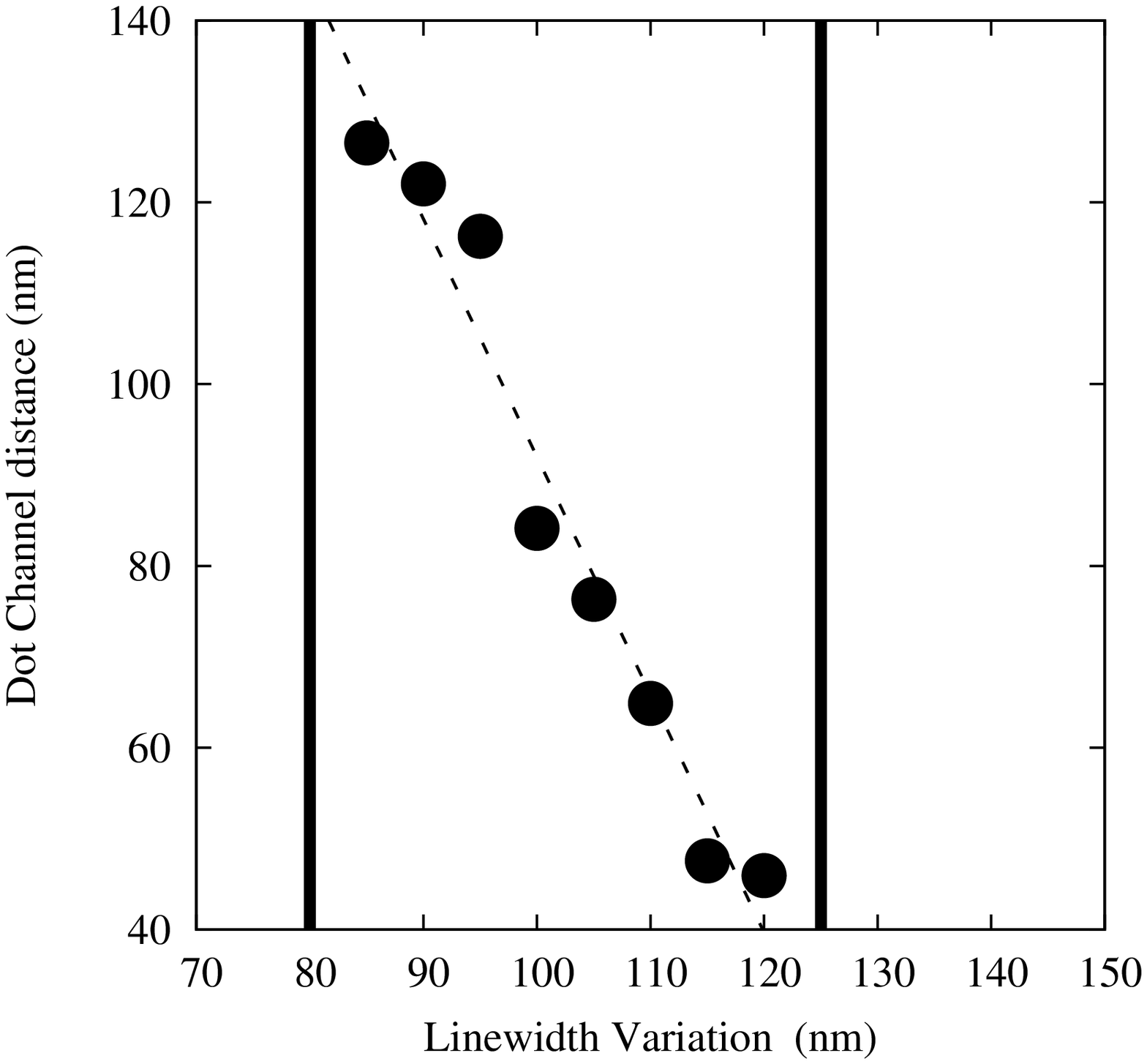}
\caption{Evolution of the main parameters (dot size, dot position, tunnel oxide thickness) between the two critical linewidths.\label{fig:param_soi}}
\end{figure}

Process simulation can help  to have an estimate of the variation  of  the dot (shape, size, position)and the parameter  range that allow for dot formation.  The dot generated by our process has a triangular shape  with a dot height often more important than the dot width. In order to simplify the discussion, the use of a  mean radius which  matches as closely as possible the dot geometry is proposed \footnote{The mean radius is extracted from the calculation of the dot surface defined by the integration of the various  finite elements.}. The variation of the dot properties as a function of the linewidth are reported Fig. \ref{fig:param_soi}. The  dot radius r$_{dot}$  decreases linearly with the initial linewidth. It can  be described by the following relationship with a small set of  parameters:

\begin{equation}
r_{dot}=0.33(d-d_{max})+r_{max} \quad  (nm)
\label{fit_radius2}
\end{equation}
\medskip

\noindent where $d$ is the linewidth, d$_{max}$ is the critical line for the dot creation (125 nm) and r$_{max}$ is the associated maximum radius (20 nm). 

Another interesting parameter is the variation of the dot position  with the nanowire initial width. In general, the dot center is located 30 nm below the top silicon surface close to the midpoint  between the top surface and the arsenic profile maximum. It can be observed that the thickness of the oxide separates the dot from the channel increases as the initial linewidth decreases (Fig. \ref{fig:param_soi}).  As clearly shown in Fig. \ref{fig:central_soi2},  the dot position  approaches 15 nm for a linewidth around 90 nm. This trend can be simply described by the following linear  empirical law as a function of the difference between the linewidth and  d$_{max}$ = 125 nm :

\begin{equation}
r_{center}=0.484 (d-d_{max})   \quad (nm)
\label{fit_dot_pos}
\end{equation}
\medskip

This effect can be explained by the fact that, for thinner structures, the oxidation  is more important in the region of  strong arsenic gradient since there is a smaller amount  of silicon to oxidize for the same thermal budget.

Finally, the evolution of the tunnel oxide thickness $t_{ox}$  (the dot-channel distance)  is also  reported. The tunnel oxide increases almost  linearly with downscaling as the thermal budget is kept constant and not adjusted for the critical linewidth study.  

\begin{equation}
t_{ox}=26.59+2.616(d-d_{max})   \quad (nm)
\label{fit_tox}
\end{equation}
\medskip

\subsection{Conclusion}

The existence of two critical linewidth (125 nm and 80 nm) for the dot existence has been clearly established by the process simulation in agreement with the experiments. The different structures are  correctly described by process simulation despite the limitations  in section III.  The difference  (15 $\%$) observed  on the critical line for the dot consumption  could be explained by the limits of the experimental technique and also  by the limitations related to strain effects at the nanometer scale, two dimensional arsenic dopant diffusion and segregation. Process simulation also provides interesting trends for dot geometry characteristics.  The mean radius of the nanofloating gate  scales down almost linearly with the linewidth. On the other hand, the dot-channel distance and the dot center position increases due to a strong oxidation rate at the bottom of the nanofloating gate. These variations have been quantified and empirical variation laws extracted.

\section{Downscaling}

One challenge faced  by alternative nanometric devices is their scalability. In addition, aggressive size reduction would open the perspective to operate close to single electron mode of operation. The main objective of this section is to evaluate the possibility of downscaling  this alternative device using the same simulation strategy that was above detailed.

\subsection{Scaling issues}
  
Downscaling of the  standard device by a factor 2 is our target. The first issue is to adapt our process for the implementation on a thinner silicon film of 100 nm. The film thickness reduction mainly impacts the implantation conditions. The objective is to  create a sharp arsenic gradient in a more localized area. First simulations show that the arsenic peak should not be too close to the  silicon surface in order to create the dot  since  oxidation takes also place under the nitride mask. Therefore, the arsenic maximum has to  be located into the top third of the silicon film as in the standard configuration. A reduction of the  implantation energy around 60 keV is therefore recommended. The scaling of the dose  must be investigated but other implantation conditions (tilt angle) could remain unchanged. The second issue is that in order to be able to form the constrictions, the region where the channel will be formed should be significantly thicker that where the dot will.

\subsection{Linewidth variation}

\begin{figure}[tbp]
\center
\includegraphics[width=10cm]{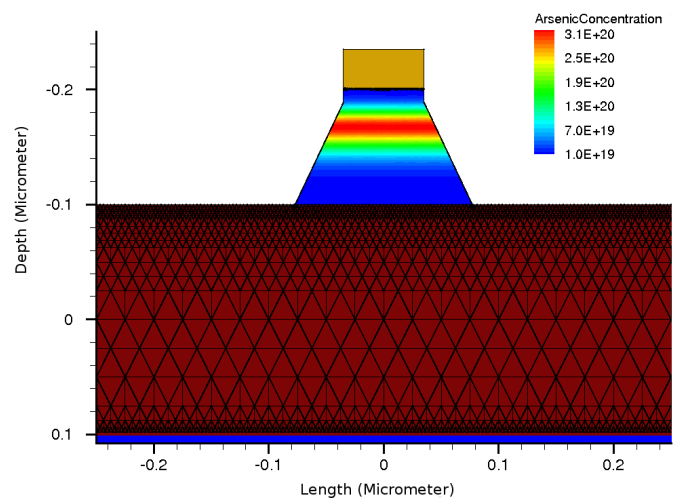}
\caption{Initial configuration (70 nm baseline) used in the downscaling study. Overetching is not considered here and a standard  anisotropic etching has been perform to  match the nanowire shape \label{fig:exp_scale}}
\end{figure}

\begin{figure}[tbp]
\center
\includegraphics[width=7cm]{120nm_bw_v2.ps}
\includegraphics[width=7cm]{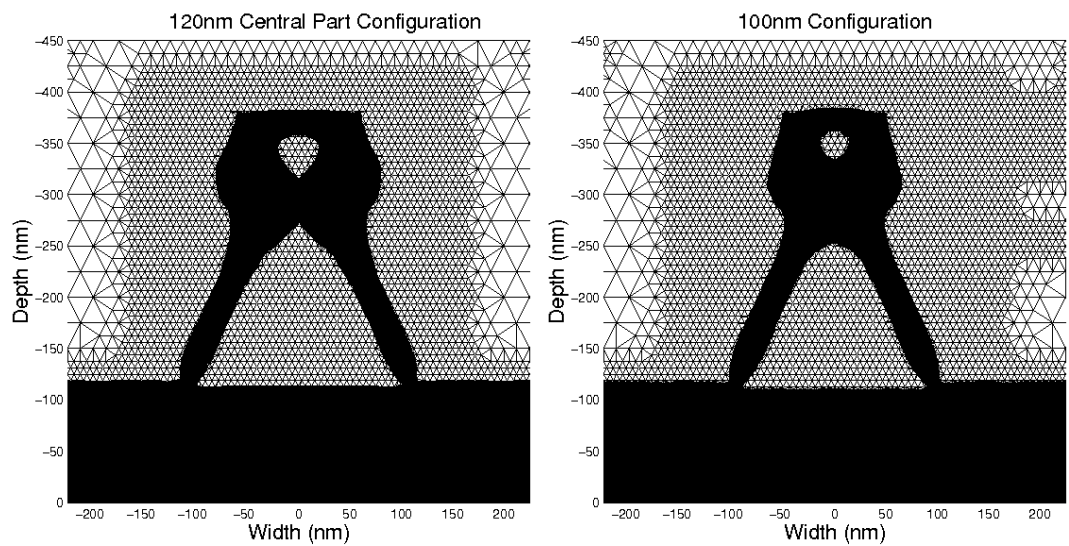}
\caption{Evolution of the central configuration for 120 nm and 100 nm linewidth. \label{fig:central_soi2}}
\end{figure}

In the downscaling study, the initial configuration shown in Fig. \ref{fig:exp_scale} has been used. Arsenic implantation has been performed with the parameters  chosen in the previous section (Energy 60 keV, Arsenic dose 1e15 at/cm$^{2}$). A nitride film has been deposited on top of the oxide layer. No overetching  is considered since  the main objective is to identify the trends. Moreover it will complexity the discussion of the results.  A standard  sidewall has been chosen to define the shape at the  bottom of the mesa-structure. 

\begin{figure}[tbp]
\begin{center}
\includegraphics[width=7cm]{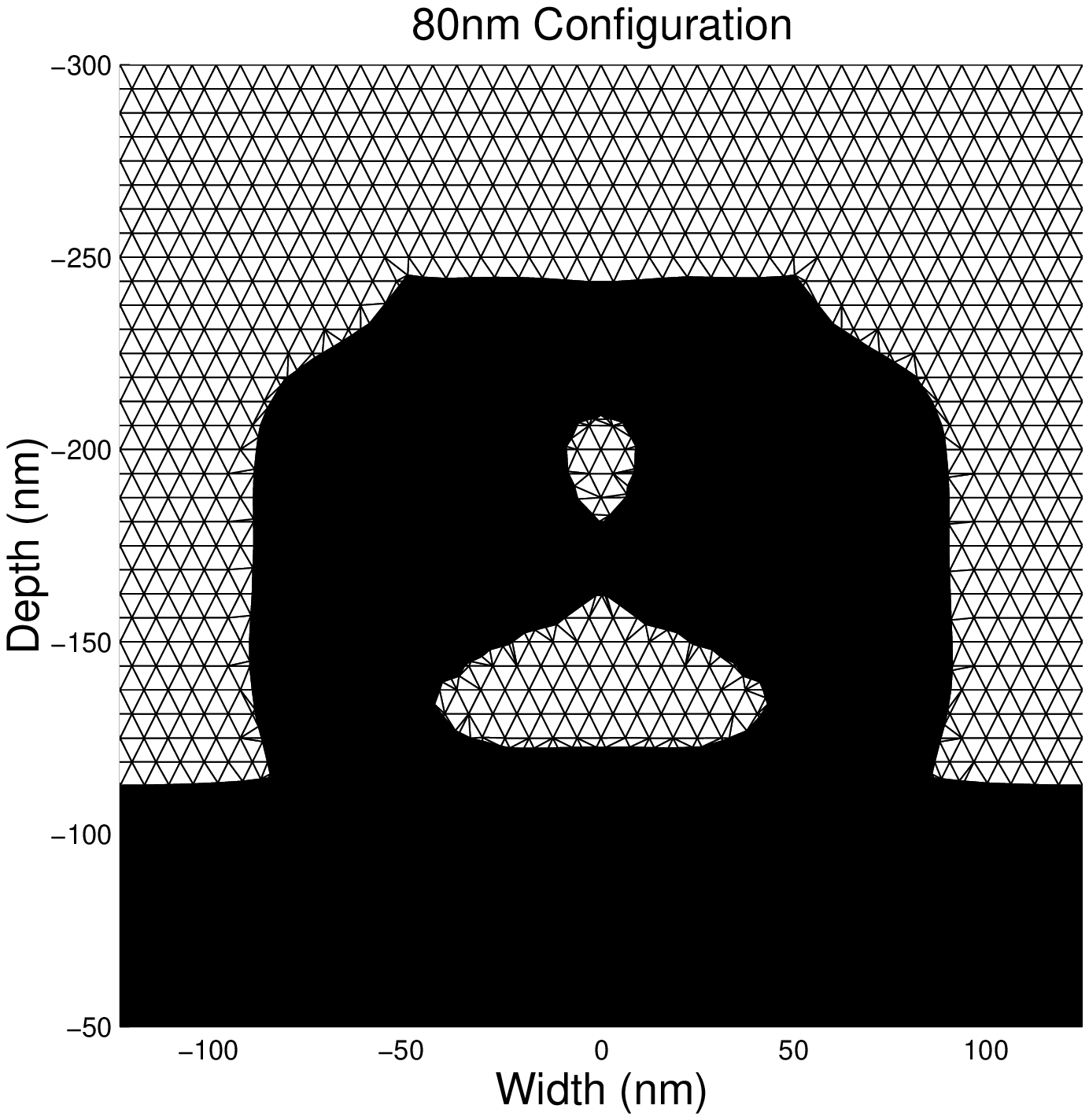}
\includegraphics[width=7cm]{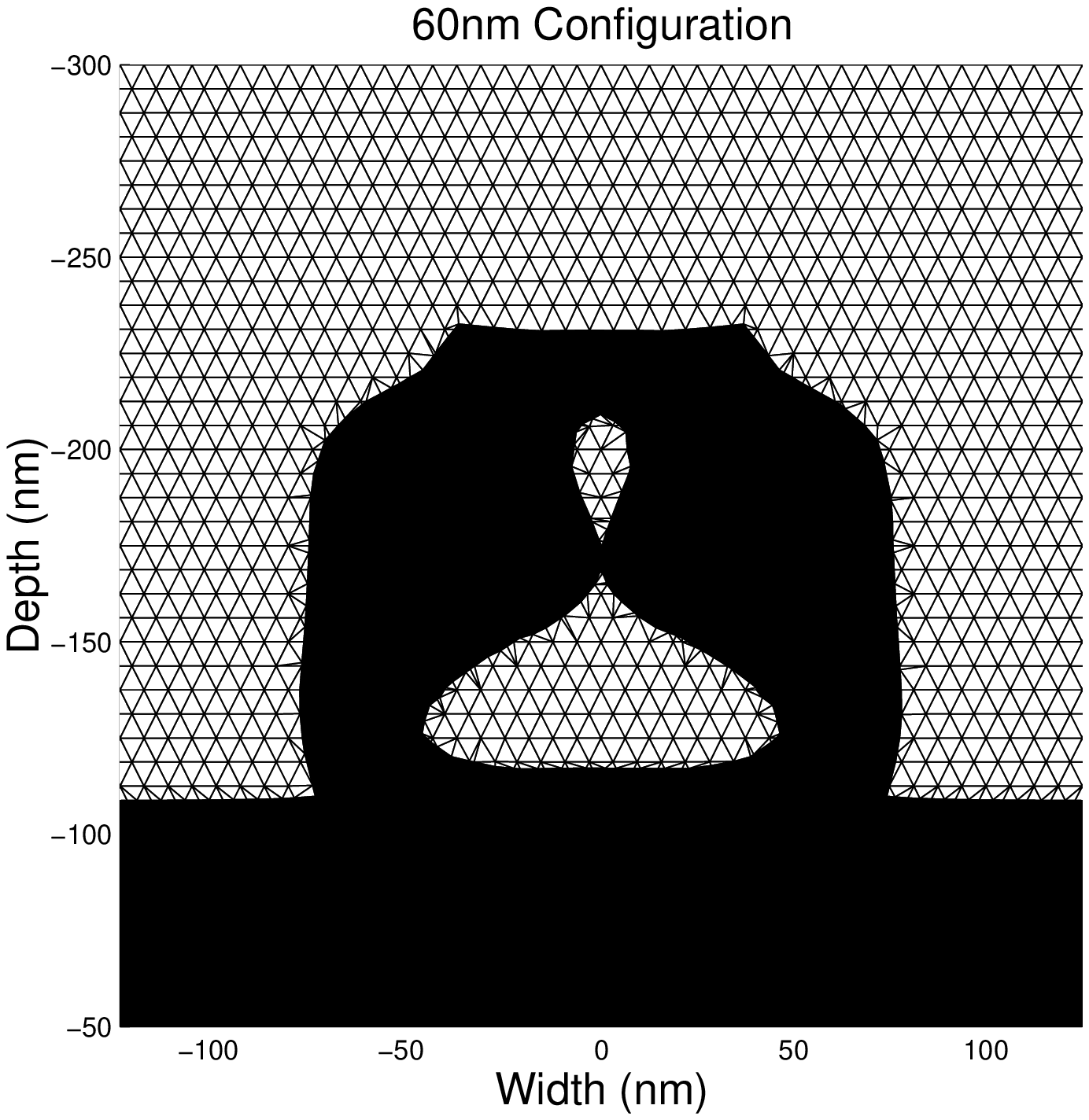}
\includegraphics[width=7cm]{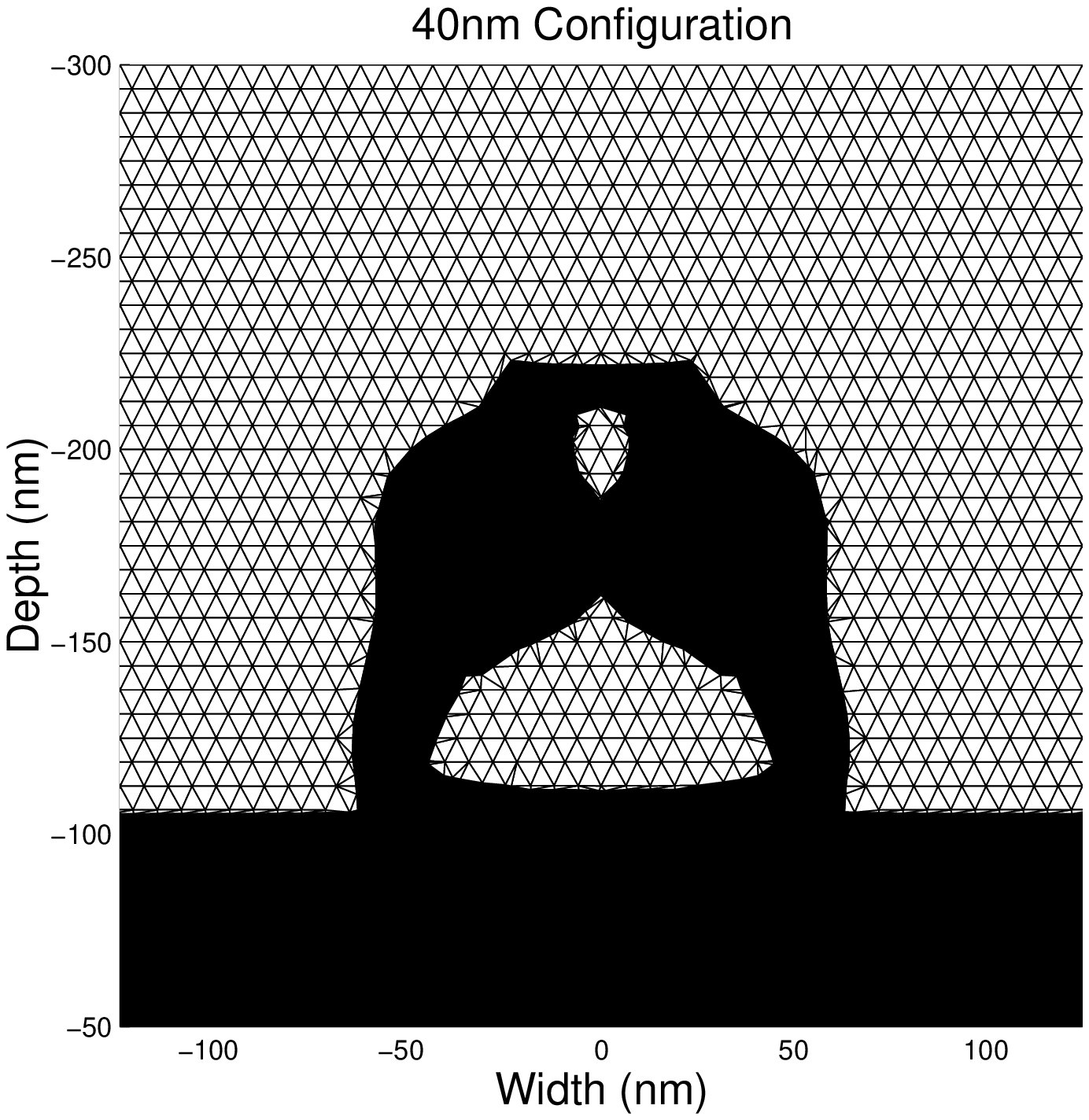}
\end{center}
\caption{Simulated 80 nm, 60 nm and 40 nm linewidth downscaled configurations. The oxidation time is set according to the estimation reported in Fig. \ref{fig:sum_param}. \label{fig:downscaled_soi}}
\end{figure}

An extensive set of simulations  has been performed as a function of  the initial linewidth. A low temperature  wet oxidation performed at (800$^{\circ}$C), the optimal temperature.  The final resulting configurations (Fig. \ref{fig:downscaled_soi}) are reported, close to the dot creation. First, large configurations (80 nm to 60 nm)  have been simulated. From the simulation standpoint, it clearly appears that downscaling is feasible. The final  structure is well organized since the dot is in the top half of the film and the constriction region in the bottom half of the device. Thinner structures have also  been investigated (from  a 50 nm configuration  down to 20 nm). Very surprisingly,  the structure dot-channel has been observed to be scalable  down to 20 nm  with a dot around 5 nm as shown in Fig. \ref{fig:sum_param}.

\subsection{Evolution of dimensional properties of the dot}

\begin{figure}[tbp]
\includegraphics[width=7cm]{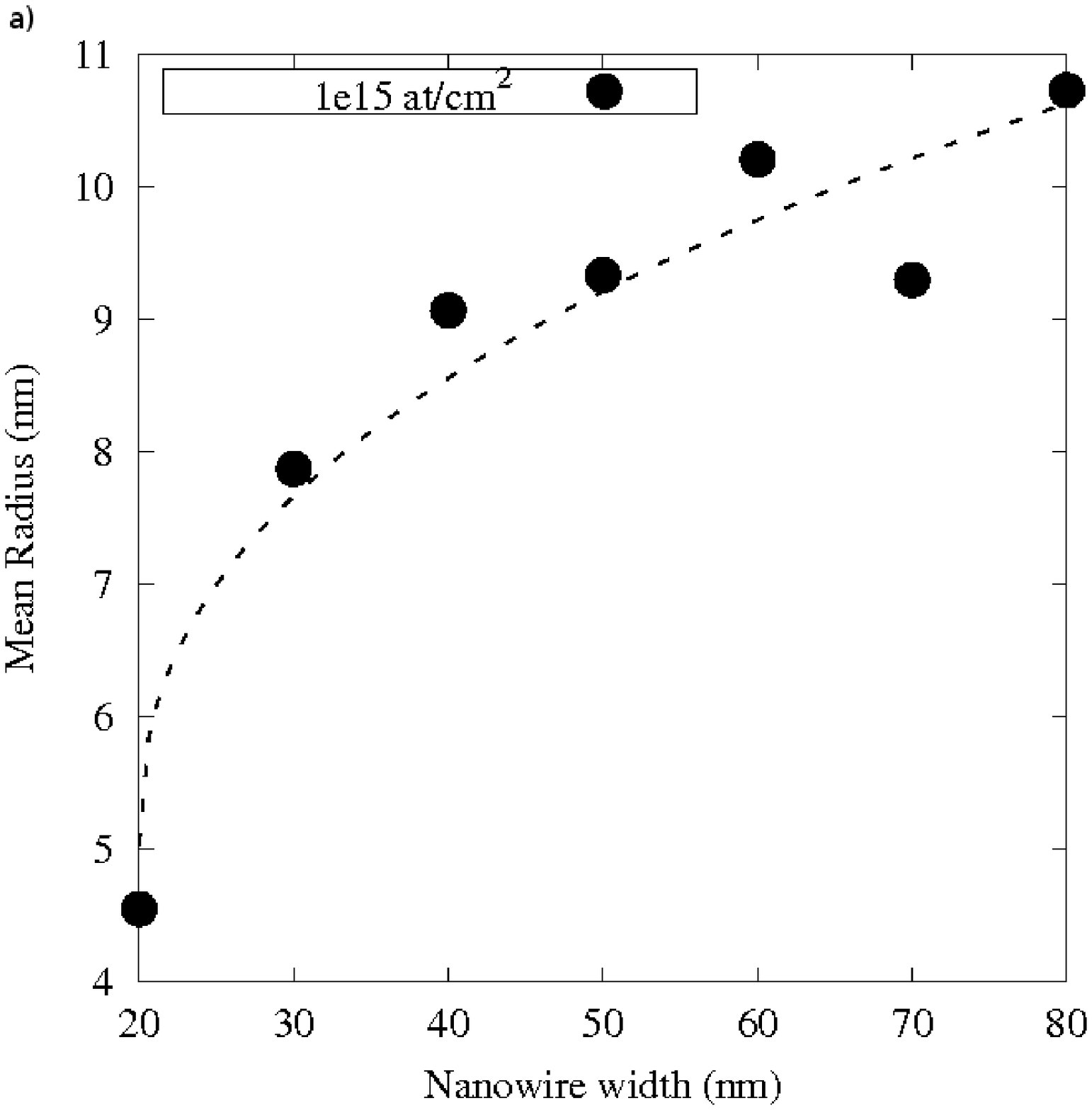}
\includegraphics[width=7cm]{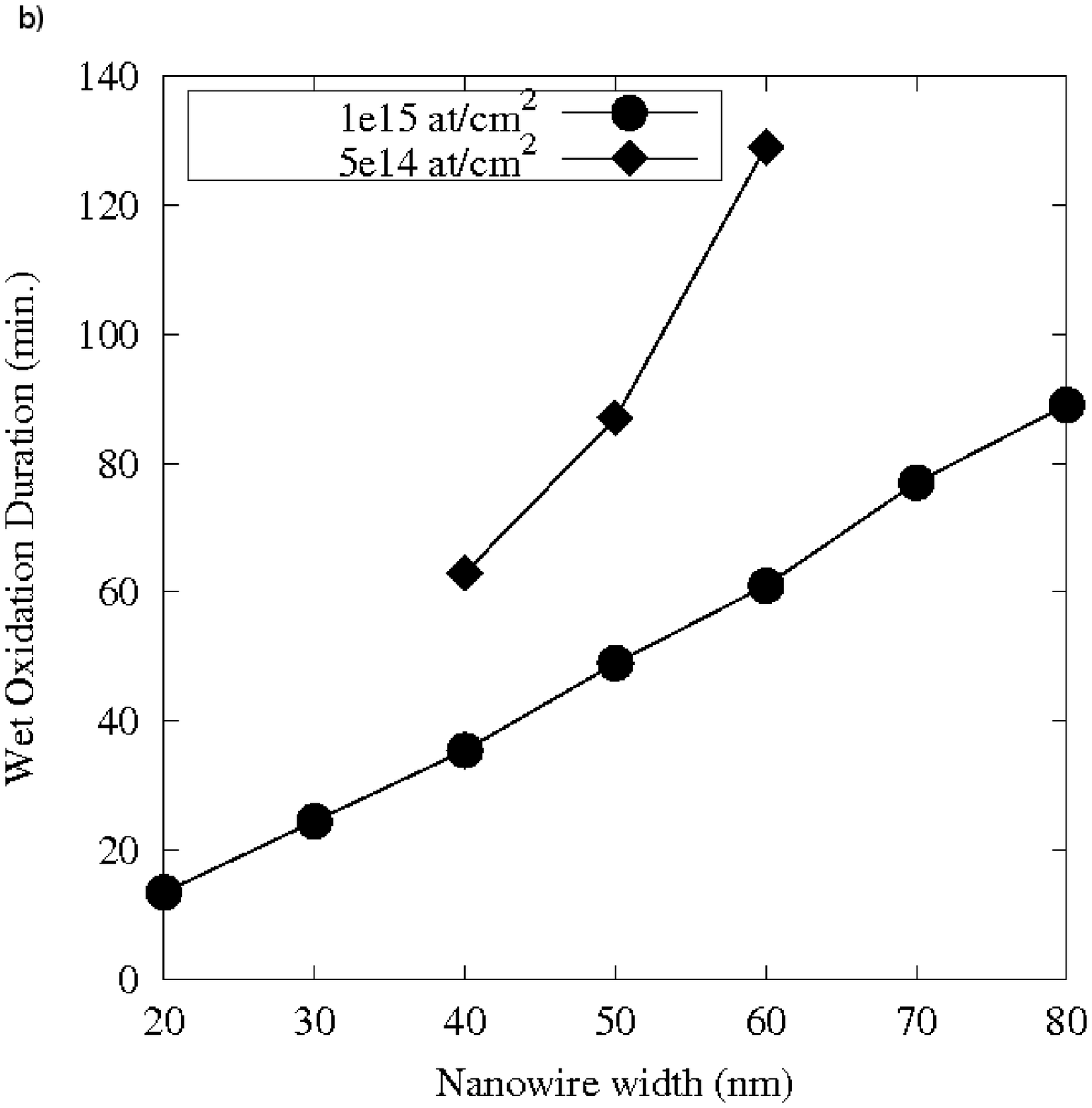}
\caption{Summary of the parameters evolution in the downscaling study. a) The dot radius evolution as a function of the initial  nanowire linewidth is presented. b) The  oxidation time necessary to create the silicon dot with decreasing linewidth for the standard dose (1$\times$ 10$^{15}$ at/cm$^{2}$) and with a lower dose (5$\times$ 10$^{14}$ at/cm$^{2}$) is reported.  It can be observed that the thermal budget downscales linearly with decreasing linewidth and that  reducing the dose by a factor two is not suitable because the range for the dot existence is clearly sharper [60-40] nm only.\label{fig:sum_param}}
\end{figure}

Top Fig. \ref{fig:sum_param} summarizes the evolution of two parameters (the dot mean radius and the thermal budget of the wet oxidation step)as a  function of the nanowire linewidth. For a large configuration of 80 nm, the mean radius approaches the 10 nm target.  As expected, the objective to reduce the size of the dot by a factor two has been achieved.  The dot radius reduces for narrower configurations down to  5 nm for a 20 nm configuration. This figure emphasizes the interest of  this approach to generate a large spectrum of dot using single parameter variation. The bottom graph in Fig.  \ref{fig:sum_param}  shows the evolution of the oxidation time at the onset of the dot creation. The various final structures presented just before  obtained by setting the oxidation duration close to the creation time for each linewidth. As shown in  Fig. \ref{fig:sum_param}, the thermal budget downscales almost linearly with decreasing linewidth. For a large configuration, the oxidation time is about 80 min. and is less than 20 min. for the narrowest device. The impact of dose reduction by a factor 2 has also been evaluated. The linewidth range for the dot existence is clearly much sharper indicating that the doping gradient should not be further reduced.

\subsection{Conclusion}

Based on the downscaling study, the fabrication of a single electron memory cell has been achieved with a reduction by a factor 2 with respect to the original design \cite{Xiaohui06}being kept constant. The silicon film thickness has been reduced to 100 nm and the implantation energy reduced to 55 keV (close to the best theoretical value of 60 keV). The formed dot memory  has been observed through the gate and has a diameter of about 5-10 nm. The downscale device proposed by simulation is  the 60 nm configuration of Fig. \ref{fig:downscaled_soi} with a dot size of 10 nm, a gate oxide of about 20 nm and a tunnel oxide of   15 nm. A preliminary  device simulation study has shown that the device can operated for  relatively thick oxides  (15 nm for the gate oxide and 35 nm the for tunnel oxide) \cite{Tang02}.  As already reported in a previous paper,  single hole effects  have been observed by several electrical characterizations at room temperature in the implementation of the downscale device \cite{Xiaohui06}. As the process space allows some process  modifications,  more work should be done  in order  to optimise the device space for this device.

\section{Conclusion}

Important conclusions can be drawn  on the two major topics. First, the concept of   single silicon dot generated by oxidation is of interest. It has been shown that a large range  of  single dots can be generated from 20 nm nanometer down to 5 nm. This method is very complementary to other  techniques for which smaller dot are often obtained. Single dots  generated with this technique hold the noteworthy  property of tight dot size control.  The proposed flash memory design and the corresponding process open new perspectives  for the integration of this concept in self-assembled nanowires to generate mass flash memory or the design of complex device with several single dots in the central part to fabricate single electron pumps \cite{Zorin}.

Secondly, on the process modeling point of view, this study is one example of the interest of process modeling to speed-up the development of nanodevices. The impact of various parameters and configurations has been investigated. Major trends and  sharp compromises can be determined. This work also illustrates the challenges  faced by  process simulation. The development of new devices needs the simulation of various coupled phenomenas (diffusion, oxidation, segregation, strain relaxation).  Atomistic simulations are often proposed as an interesting alternative to simulate the process fabrication of such devices. But as clearly shown by this study, the main issue is  to tackle the simulation of process fabrication  and the interaction of  relatively large silicon pattern with nanometric objects. Improving the limitations of continuous models  and developing  multi-scale approaches remain mandatory.

\section*{Acknowledgment}
This work was supported by the European Union  under Contract IST-2001-32674 SASEM Project (Self-Aligned Single Electron Memory and Circuits). One of the authors (C. K.) would like to thanks the reviewers of the SASEM project, Professor Asen Asenov (University of Glasgow) and Professor Niemeyer (PTB) for giving very fruitful  directions during the project. The support of Patrick Van Hove from the European Commission (Future and Emerging Technologies) was also appreciated \cite{Van-Hove06}. This work benefits also from fruitful discussions  with E. Lampin (IEMN) on implantation and diffusion simulations,  with  V. Senez (IEMN) on oxidation and mechanical strain  and  with C. Bonafos (CEMES-Toulouse) on silicon nanocrystals oxidation kinetics. The regular support of Jean-Michel Droulez on  workstations, software, licence, script for this project is also acknowledged.

\newpage

\end{document}